\documentclass{article}

\newlength{\abstractwidth}
\usepackage{cancel}
\usepackage{hyperref}
\usepackage{color}
\usepackage{graphicx}

\usepackage{tikz, pgf}
\usepackage{tkz-fct}
\usepackage{pgfplots}
\usetikzlibrary{shapes.misc}
\usetikzlibrary{shapes,snakes}
\usetikzlibrary{decorations.pathmorphing}	
\usetikzlibrary{decorations.markings}

\usetikzlibrary{arrows.meta}

\newcommand{\backmidarrow}{\tikz  \draw[thick,-{Straight Barb[angle'=60,scale=1.5]}]  ((.1,0)-- (0,0);}
\newcommand{\midarrow}{\tikz  \draw[thick,-{Straight Barb[angle'=60,scale=1.5]}]  (0,0) -- (.1,0);}

\usepackage{amsmath}
\usepackage{amssymb}
\usepackage{latexsym}
 \usepackage{setspace}
 
 \usepackage{caption}
\usepackage{subcaption}

\numberwithin{equation}{section}

\renewcommand{\thefootnote}{\fnsymbol{footnote}}
\renewcommand{\thanks}[1]{\footnote{#1}}
\newcommand{\starttext}{
\setcounter{footnote}{0}
\renewcommand{\thefootnote}{\arabic{footnote}}}
\newcommand{\bea}{\begin{eqnarray}}
\newcommand{\eea}{\end{eqnarray}}
\newcommand{\be}{\begin{eqnarray}}
\newcommand{\ee}{\end{eqnarray}}

\def\ie{\begin{equation}\begin{aligned}}
\def\fe{\end{aligned}\end{equation}}
\def\half{{\scriptstyle \frac 12}}

\def\threeh{{\scriptstyle \frac 32}}
\def\fiveh{{\scriptstyle \frac 52}}

\def\ie{\begin{equation}\begin{aligned}}
\def\fe{\end{aligned}\end{equation}}
\def\cB{{B}}
\def\cC{{\cal C}}
\def\cD{{\cal D}}

\def\cG{{\cal C}}

\def\cN{{\cal N}}

\def\Z{{\mathbb Z}}

\def\nn{\nonumber}

\def\C{\cC}

\def\tN {\tilde N}

\begin{document}

\starttext

\setcounter{footnote}{0}

\begin{flushright}
{\small QMUL-PH-22-30}
\end{flushright}

\vskip 0.3in

\begin{center}

\centerline{\large \bf  Modular-invariant large-$N$ completion  of an integrated correlator}
\centerline{\large \bf  in $\mathcal{N}=4$ supersymmetric Yang--Mills theory} 

\vskip 0.2in

{ Daniele Dorigoni$^{(a)}$, Michael B. Green$^{(b)}$, Congkao Wen$^{(c)}$, and Haitian Xie$^{(c)}$} 
   
\vskip 0.15in

{\small ($a$) Centre for Particle Theory \& Department of Mathematical Sciences, 
}\\
\small{Durham University, Lower Mountjoy, Stockton Road, Durham DH1 3LE, UK}

\vskip 0.1in

{ \small ($b$) Department of Applied Mathematics and Theoretical Physics }\\
{\small  Wilberforce Road, Cambridge CB3 0WA, UK}

\vskip 0.1in

{\small  ($c$) Centre for Theoretical Physics, Department of Physics and Astronomy,  }\\ 
{\small Queen Mary University of London,  London, E1 4NS, UK}

\vskip 0.15in

{\tt \small  daniele.dorigoni@durham.ac.uk,  mbg15@cam.ac.uk,  c.wen@qmul.ac.uk, h.xie@se21.qmul.ac.uk}

\vskip 0.5in

\begin{abstract}
 
 \vskip 0.1in

The use of supersymmetric localisation has recently led to modular covariant expressions for certain
integrated correlators of half-BPS operators in $\mathcal{N} = 4$ supersymmetric Yang--Mills theory with a general
classical gauge group $G_N$. Here we determine generating functions that encode such integrated correlators
for any classical gauge group and provide a proof of previously conjectured formulae. This gives a systematic understanding of the relation between
properties of these correlators at finite $N$ and their expansions at large $N$. In particular, it determines
a duality-invariant non-perturbative completion of the large-$N$ expansion in terms of a sum of novel
non-holomorphic modular functions. These functions are exponentially suppressed at large
$N$ and have the form of a sum of contributions from coincident $(p, q)$-string world-sheet instantons.

\end{abstract}                                            
\end{center}

\baselineskip=15pt
\setcounter{footnote}{0}

\newpage

\setcounter{page}{1}
\tableofcontents

\newpage

\section{Introduction and outline}

The exact expressions for certain integrated correlators of four superconformal primary operators of the stress tensor multiples in $\cN=4$ supersymmetric Yang--Mills (SYM) theory with gauge group $G_N$  can be determined by localisation in terms of the partition function,  $Z_{G_N}(\tau,\bar\tau,m)$, of $\cN=2^*$  SYM that was derived  by Pestun \cite{Pestun:2007rz} in terms of the Nekrasov partition function \cite{Nekrasov:2002qd}. The $\cN=2^*$ theory reduces to the $\cN=4$ theory in the limit $m\to 0$, where  the parameter $m$ is the hypermuitiplet mass and $Z_{G_N}(\tau,\bar\tau,m)|_{m=0}=1$. In  \cite{Binder:2019jwn} the expression 
\bea
\cC_{G_N}(\tau,\bar\tau):=\frac{1}{4} \Delta_\tau\partial_m^2 \log Z_{G_N}(\tau,\bar\tau,m) |_{m=0}\,,
\label{eq:firstcorr}
\eea
was argued to be proportional to an integrated correlator of four superconformal stress-tensor primaries with a specific integration measure.  The quantity $\Delta_\tau := 4 \tau_2^2\partial_\tau \partial_{\bar \tau}$ is the laplacian on the hyperbolic plane parametrised by the coupling constant $\tau= \tau_1+i\tau_2 := \frac{\theta}{2\pi} + i \frac{4\pi}{g_{_{Y\!M}}^2}$, with $\theta$ the theta angle and $g_{_{Y\!M}}$ the Yang--Mills coupling constant.

 The first few terms in the large-$N$ expansion of $\cC_{G_N}(\tau,\bar\tau)$   in the 't Hooft limit (in which $\lambda = g_{_{Y\!M}}^2 N$ is fixed) for $SU(N)$ gauge group were studied in an expansion in powers of $1/\lambda$  in  \cite{Binder:2019jwn} and similarly for general classical groups in \cite{Alday:2021vfb}.    As shown in \cite{Chester:2019jas}, the  coefficients in the perturbative  $1/N$ expansion at fixed $\tau$ are modular functions that make  $SL(2,\Z)$ Montonen--Olive duality \cite{Montonen:1977sn}  (also known as S-duality) manifest.  These coefficients are sums of non-holomorphic Eisenstein series with half-integer index. The perturbative pieces of the correlator can be extracted relatively easily  from the localised expression for $Z_{G_N}(\tau,\bar\tau,m)$    for any value\footnote{It was shown in \cite{Wen:2022oky} that the perturbative contribution of integrated correlators has an interesting intepretation in terms of periods of certain conformal Feynman integrals and agrees with explicit perturbative computations \cite{Gonzalez-Rey:1998wyj,Eden:1998hh,Bianchi:2000hn,Eden:2000mv,Eden:2011we,Bourjaily:2015bpz,Bourjaily:2016evz} up to four loops.}  of $N$ \cite{Chester:2019pvm, Chester:2020dja, Alday:2021vfb}, but extracting  the explicit instanton contributions that are contained in the Nekrasov partition function is more  involved. 
 
 However, in \cite{Dorigoni:2021bvj,Dorigoni:2021guq, Dorigoni:2022zcr}  a novel expression for the integrated correlator was proposed that is valid for any classical gauge group $G_N$ and finite $\tau$.\footnote{See \cite{Dorigoni:2022iem} for a recent review, as well as \cite{Green:2020eyj, Dorigoni:2021rdo} for the extension to higher-point maximal $U(1)_Y$-violating correlators and \cite{Paul:2022piq} for the generalisation to integrated four-point correlators involving operators with higher conformal weights using $SL(2, \mathbb{Z})$ spectral theory.}  This takes the form  of a double lattice sum,
 \bea
\cC_{G_N} (\tau,\bar\tau)  = 
 \sum_{(m,n)\in\mathbb{Z}^2}   \int_0^\infty  \left[ e^{ - t\,  Y_{mn}(\tau, \bar \tau)} B^1_{G_N}(t) + e^{ - t \, Y_{mn}(2\tau, 2\bar \tau)} B^2_{G_N}(t)  \right]  dt   \, ,
\label{gsun}
\eea
where we have defined the quantity
\begin{equation}
Y_{mn}(\tau, \bar \tau) :=\pi \frac{|m+n\tau|^2}{ \tau_2}\,.\label{eq:Ymn}
\end{equation} 
The coefficient functions  $B^1_{G_N}(t)$ and $B^2_{G_N}(t)$ are rational functions of the following form, 
 \begin{equation}
B^i_{G_N}(t)= \frac{\mathcal{Q}^i_{G_N}(t)}{(t+1)^{n^i_{G_N}}}\,,\label{eq:BN}
 \end{equation}
 where $i=1,2$, $n^i_{G_N}$ is an integer and $\mathcal{Q}^i_{G_N}(t)$ is a degree $n^i_{G_N}\!\!-2$ polynomial with the ``palindromic'' property $\mathcal{Q}^i_{G_N}(t) =t^{n^i_{G_N} \!-1} \mathcal{Q}^i_{G_N}(t^{-1})$.
 
  For simply-laced groups $G_N = SU(N)$, $SO(2N)$  the correlators are expected to be invariant under the $SL(2,\Z)$ action $\tau \to \gamma\cdot \tau = \frac{a\tau+b}{c\tau+d}$ with $\gamma = \left(\begin{smallmatrix}
a& b \\ c & d \end{smallmatrix}\right)\in SL(2,\Z)$, which is a consequence of Montonen--Olive duality.
  In these cases, $B^2_{G_N}(t)=0$ and only $B^1_{G_N}(t)$ is non-trivial,\footnote{To simplify the notation, for these cases we will simply drop the superscript ``1'' and write $B^1_{G_N}(t) = B_{G_N}(t)$ when discussing $G_N=SU(N),SO(2N)$.} and \eqref{gsun} is manifestly invariant under $SL(2,\Z)$. For the non simply-laced classical groups $G_N = USp(2N)$, $SO(2N+1)$  \eqref{gsun} is only invariant under the congruence subgroup $\Gamma_0(2)\subset SL(2,\mathbb{Z})$.\footnote{The elements of the congruence subgroup $\Gamma_0(2)$ are given by $\gamma = \left(\begin{smallmatrix}
a& b \\ c & d \end{smallmatrix}\right)\in SL(2,\Z)$ with $c\equiv 0\,(\mbox{mod}\,2)$.} Furthermore $B^i_{G_N}(t)$ obeys the following relations
 \ie \label{eq:GNO}
 B^1_{USp(2N)}(t) =  B^2_{SO(2N+1)}(t)\,  ,\qquad  B^2_{USp(2N)}(t) =  B^1_{SO(2N+1)}(t) \, , 
 \fe
 which make Goddard--Nuyts--Olive (GNO) duality \cite{Goddard:1976qe} of \eqref{gsun}  manifest.  
 
 For example, for the $SU(N)$ theory, it was conjectured that \cite{Dorigoni:2021bvj,Dorigoni:2021guq} 
 \ie \label{eq:BSUN}
 B_{SU(N)}(t)= \frac{\mathcal{Q}_{SU(N)}(t)}{(t+1)^{2N+1}}\, , 
 \fe  and 
\begin{align}
 \mathcal{Q}_{SU(N)}(t) &\nn = -\frac{1}{4}N(N-1)(1-t)^{N-1}(1+t)^{N+1}\\
 &\left\lbrace [3+(8N+3t-6)t] P_N^{(1,-2)}\Big(\frac{1+t^2}{1-t^2}\Big) +\frac{1}{1+t} (3t^2-8 N t-3 )P_{N}^{(1,-1)}\Big(\frac{1+t^2}{1-t^2}\Big)\right\rbrace\,,
 \label{eq:Bndef}
 \end{align}
 expressed in terms of Jacobi polynomials $P_n^{(a,b)}(x)$. 
  The integrated correlator satisfies a Laplace difference equation that takes the form 
\begin{align} 
\Delta_\tau \C_{SU(N)}(\tau,\bar{\tau})   -4c_{SU(N)} & \Big[\C_{SU(N+1)}(\tau,\bar{\tau})- 2\,\C_{SU(N)}(\tau,\bar{\tau})+\C_{SU(N-1)}(\tau,\bar{\tau})\Big]\nn\\
  &-(N+1)\,\C_{SU(N-1)}(\tau,\bar{\tau}) + (N-1)\, \C_{SU(N+1)}(\tau,\bar{\tau})=0 \, ,
\label{lapdiffSUN}
\end{align}
where $\Delta_\tau=\tau_2^2 (\partial_{\tau_1}^2+\partial_{\tau_2}^2)$ is the $SL(2,\Z)$-invariant hyperbolic laplacian and $c_{SU(N)} =(N^2-1)/4$ is the central charge.       
Upon iteration, this equation  relates the integrated correlator for the theory with gauge group  $SU(N)$ to the integrated correlator for the $SU(2)$ theory (with the boundary condition $\cC_{SU(1)}(\tau,\bar\tau)=0$). Similar Laplace difference equations were also obtained for the integrated correlators with general classical gauge group $G_N$ \cite{Dorigoni:2022zcr}, with the result that all $\C_{G_N}(\tau,\bar{\tau})$ are determined in terms of $\C_{SU(2)}(\tau,\bar{\tau})$.

While it is much easier to analyse  the  dependence of the integrated correlator  on the parameters $\tau$ and $N$ starting from \eqref{gsun} than from the original expression \eqref{eq:firstcorr}, the dependence on $N$ is not transparent.  This will be remedied in the present paper, in which we will take the further step of introducing a generating function for the $N$-dependence.  
This generating function  is defined as
  \bea
 \label{eq:gendef}
 \cG_{G}(z;\tau,\bar\tau) := \sum_{N=1}^\infty   \cC_{G_N}(\tau,\bar\tau)\,z^N\,,
 \eea
 where the subscript $G$ indicates that this generates the integrated correlator for the $G_N$ gauge group for all values of $N$.
The expression \eqref{eq:gendef}  may be inverted to give
\bea
\label{eq:geninv}  
 \cC_{G_N}(\tau,\bar\tau)\  = \oint_C \frac{\cG_{G}(z;  \tau,\bar\tau)}{z^{N+1}}  \frac{dz}{2\pi i } \,,
\eea
 where $C$ denotes a contour encircling the pole at $z=0$ in an anti-clockwise direction and not encircling other singularities.
  From \eqref{gsun} we can equivalently define the generating functions for  the rational functions $B^i_{G_N}(t)$
 \bea
 \label{eq:gztdef}
\cB^i_{G}(z;t) := \sum_{N=1}^\infty   B^i_{G_N}(t)\,z^N\, ,
 \eea
 and hence introduce
\ie
 \cG^1_G(z;\tau,\bar\tau) &:= \sum_{(m,n)\in\mathbb{Z}^2}   \int_0^\infty   e^{ - t\,  Y_{mn}(\tau, \bar \tau)} B^1_{G}(z;t) dt\,,\cr
   \cG^2_G(z;\tau,\bar\tau) &:=  \sum_{(m,n)\in\mathbb{Z}^2}   \int_0^\infty  e^{ - t \, Y_{mn}(2\tau, 2\bar \tau)} B^2_{G}(z;t)    dt\,.
\fe
One of the advantages of introducing a generating function such as $\cG^i_G(z;  \tau,\bar\tau)$ is that it has a much simpler form than $\cC^i_{G_N}(\tau,\bar\tau)$. This makes $\cG^i_G(z;  \tau,\bar\tau)$ extremely convenient for analysing the large-$N$ properties of the integrated correlators.   
 
In section \ref{sec:genallN} we will determine the generating function  $\cB_{SU}(z;t)$ (which generates $B_{SU(N)}(t)$ for all $N$) by relating the integrated correlator to hermitian matrix model integrals and, in particular, this will lead to a proof of the previously conjectured expression \eqref{eq:BSUN}. The proof relies on deriving the Laplace difference equation \eqref{lapdiffSUN} from the hermitian matrix model and utilising the explicit result of the $SU(2)$ correlator in \cite{Dorigoni:2021guq}, which is the initial condition for the recursion relation \eqref{lapdiffSUN}.
Furthermore, we will show that the generating function $\cB_{SU}(z;t)$ satisfies a second order partial differential equation which leads to the Laplace difference equation \eqref{lapdiffSUN}.  The generating function will streamline the analysis of properties of the integrated correlator in different regions of parameter space by distorting the integration contour $C$ in different ways.   Some relevant properties of the hermitian matrix model and its connection to the integrated correlator with $SU(N)$ gauge group are summarised in appendix \ref{app:matrix1}.

As will be demonstrated in section~\ref{sec:largeN}, the  generating functions $\cG_{SU} (z;\tau,\bar\tau)$ and $\cB_{SU}(z;t)$  lead to an efficient procedure for determining the large-$N$ behaviour of $\cG_{SU(N)}(\tau,\bar\tau)$. This is not only a more efficient procedure for determining results that were previously derived in \cite{Dorigoni:2021guq} but also leads to new results.  We will see that the large-$N$ expansion consists of three pieces.
The first is a term proportional to $N^2$ with a constant coefficient.  The second piece is an infinite  power series in half-integer powers of $1/N$ with coefficients that are sums of half-integer non-holomorphic Eisenstein series that depend on $\tau,\bar\tau$.  These two pieces 
simply reproduce the previously determined behaviour of the integrated correlator. 

The third novel piece is non-perturbative in $N$ in the large-$N$ limit and has a leading term proportional to the modular invariant function 
\ie 
N^{2}\sum_{(m,n)\neq (0,0)}  \exp\Big(- 4 \sqrt{N  Y_{mn}(\tau, \bar \tau)}\Big)  =  N^{2} \sum_{\ell=1}^\infty \sum_{{\rm gcd} (p,q)=1 } \exp\Big(-4 \ell \sqrt {\frac{N \pi}{\tau_2}} |p+q\tau|\Big)    \, . 
\label{eq:pqstrings}
\fe

Whereas the power behaved terms in the $1/N$ expansion holographically correspond to the $\alpha'$-expansion of type IIB string amplitudes in $AdS_5 \times S^5$, the exponentially suppressed terms displayed in the above equation are  related to a sum of  $(p,q)$-string instantons (i.e.~euclidean  $(p,q)$-string world-sheets wrapping a two dimensional 
manifold).  
The complete non-perturbative contribution with leading behaviour  \eqref{eq:pqstrings} is given by a sum of new non-holomorphic modular functions, $D_N(s;\tau,\bar\tau)$, which are generalisations of non-holomorphic Eisenstein series that are exponentially suppressed at large $N$ and fixed $\tau$.  Some properties of these functions are discussed in appendix~\ref{app:newmod}.

The formal sum of the asymptotic power series expansion in half-integer powers of $1/N$  and these novel non-perturbative terms provides the complete large-$N$ transseries expansion of the integrated correlator $\cG_{SU(N)}(\tau,\bar\tau)$.  The Borel-Ecalle resummation of this transseries produces a well-defined and unambiguous analytic continuation for all values of $N$.  In particular it coincides with the finite $N\in \mathbb{N}$ results.

The presence of a third, non-perturbative, piece was previously arrived at in the 't Hooft limit, in which  $N\to \infty$  with  $\lambda=g_{_{Y\!M}}^2N$  fixed,  by use of a resurgence argument based on the non-summability of the large-$\lambda$ expansion \cite{Dorigoni:2021guq, Collier:2022emf, Hatsuda:2022enx}.
In section \ref{sec:tHooft} we demonstrate how the $SL(2,\Z)$-invariant expression for the non-perturbative contribution reduces to such an expression in a suitable limit.
We also consider the same non-perturbative terms in the regimes $\lambda = O(N^2)$, or $\lambda = O(1)$, where a different picture emerges and we retrieve the large-$N$ non-perturbative expansions obtained in \cite{Hatsuda:2022enx} by exploiting resurgence arguments for the large genus behaviour of the perturbative genus expansion.
 When $\lambda=O(N^2)$  the non-perturbative terms take a form that resembles the effects of electric $D3$-branes that arise in \cite{Drukker:2005kx}  in  the holographic description  of Wilson loops.   When $\lambda = O(1)$ the non-perturbative contributions resemble the magnetic $D3$-branes also discussed in that reference.
 Details of these expressions will be given in  appendix \ref{app:zeromode}.

The generating functions for the integrated correlators with gauge group $SO(n)$ are derived in section  \ref{sec:gengroup}, and analogous large-$N$ properties are found for these more general gauge groups.  In particular, the integrated correlator for the theory with simply-laced group $SO(2N)$, again receives  $(p,q)$-string instanton corrections. For the non simply-laced gauge group $SO(2N+1)$ there are not only $SL(2,\Z)$-invariant contributions from  $(p,q)$-string world-sheet instantons, but also $\Gamma_0(2)$-invariant contributions from   $(p,2q)$-string instantons.  This restriction is due to the fact that $\mathcal{N}=4$ SYM with gauge group $SO(2N+1)$ is only S-duality invariant  under a congruence subgroup of $SL(2, \mathbb{Z})$,  namely $\Gamma_0(2)$. 
 The same statements apply to the integrated correlator with gauge group  $USp(2N)$, which is related to the $SO(2N+1)$ case by GNO duality.  A detailed description of the derivation of  the generating function for the integrated correlator with gauge group $SO(n)$ is given in appendix \ref{app:clagroup}.

   \section{A generating function for all $SU(N)$}
   \label{sec:genallN}
   
   The form of the function $B_{SU(N)}(t)$ given in \eqref{eq:BSUN}  and  \eqref{eq:Bndef} was conjectured  in \cite{Dorigoni:2021guq, Dorigoni:2022zcr}  based on the analysis of the perturbative part of $\cC_{SU(N)}(\tau,\bar\tau)$   \eqref{eq:firstcorr}   and the explicit evaluation of a variety of non-perturbative  instanton contributions for a wide range of values of $N$.  In this section we will first determine the generating function for $SU(N)$ starting from the conjectural functional form  \eqref{eq:BSUN}-\eqref{eq:Bndef} of  $B_{SU(N)}(t)$. We will then prove that the same generating function can be derived from properties of correlation functions in $N\!\times\!N$ hermitian matrix models.
 
 To begin with we will  determine various properties of the relevant generating functions.
 The function  $\cC_{SU}(z;\tau,\bar\tau)$  \eqref{eq:gendef} can be obtained by substituting the expression  for  $B_{SU(N)}(t)$ \eqref{eq:Bndef} into $\cB_{SU}(z;t) := \sum_{N=1}^\infty B_{SU(N)}(t) z^N$ and making  use of the generating function for  Jacobi polynomials,
 \begin{equation}
 \sum_{n=0}^{\infty} P_n^{(a,b)}(x) z^n  = 2^{a+b} R^{-1} (1-z +R)^{-a} (1+z+R)^{-b}\,,
 \end{equation}
 where $R= \sqrt{1-2x z+z^2}$. 
The result is the generating function for $B_{SU(N)}(t)$  \eqref{eq:BN},
 \begin{equation}
 \cB_{SU}(z;t) := \sum_{N=1}^\infty B_{SU(N)}(t) z^N = \frac{3 t z^2 \left[(t-3) (3 t-1)(t+1)^2 -
   z(t+3) (3 t+1) (t-1)^2 \right]}{2 (1-z)^{\frac{3}{2}}
   \left[(t+1)^2-(t-1)^2 z\right]^{\frac{7}{2}}} \, ,
   \label{eq:gresult}
  \end{equation}
which leads to
  \begin{equation} \label{eq:cGSU}
  \cG_{SU}(z;\tau,\bar\tau) := \sum_{(m,n)\in\Z^2} \int_0^\infty e^{-t \, Y_{mn}(\tau, \bar \tau)}  \cB_{SU}(z;t)  dt\,.
  \end{equation}
 For future reference we notice that the generating function \eqref{eq:gresult} has a branch-cut located on the interval $z\in \big[1,\frac{(t+1)^2}{(t-1)^2} \big]$. 
  
 This generating function satisfies several properties of note,
 \ie
 \cB_{SU}(z;t) = t^{-1} \cB_{SU}(z; t^{-1})\,, \qquad\qquad 
 \cB_{SU}(z;t) =-  \cB_{SU}(z^{-1}; -t)  \,,\nn 
 \fe
\ie
 \int_0^\infty \frac{ \cB_{SU}(z;t)}{\sqrt{t}} dt = 0\,,\qquad\qquad
 \int_0^\infty  \cB_{SU}(z;t) dt =  \frac{z^2}{4(1-z)^3} = \sum_{N=1}^\infty \frac{N(N-1)}{8} z^N\,.
 \label{eq:propGSU2}
 \fe
 The first of these equations is an inversion relation that follows automatically from the lattice sum definition of the integrated correlator \eqref{gsun}, as was pointed out in \cite{Collier:2022emf} where the lattice sum was re-expressed in terms of a modular invariant spectral representation.
The second equation in \eqref{eq:propGSU2}   is an inversion relation in the variable $z$ which relates the $SU(N)$ correlator with coupling $g_{_{Y\!M}}^2$ to the $SU(-N)$ correlator with coupling $-g_{_{Y\!M}}^2$, as previously discussed in \cite{Dorigoni:2022zcr}.
 
 The Laplace difference equation \eqref{lapdiffSUN} satisfied by the integrated correlator $  \cG_{SU(N)}(\tau,\bar\tau) $ translates into a partial differential equation in $(z,t)$ for $\cB_{SU}(z;t)$,  
  \bea 
  t \partial_t^2 ( t \cB_{SU}(z;t))  = 2 (z^{-1}\!+1+z)\cB_{SU}(z;t)+2(z-1)(2z+1) \partial_z \cB_{SU}(z;t)+(z-1)^2 z \partial_z^2 \cB_{SU}(z;t)\,,\label{eq:lapdiffgsu}
  \eea
  or equivalently, in $(z,\tau,\bar\tau)$ for $\cG_{SU}(z;\tau,\bar\tau)$
\bea
 \Delta_\tau \cG_{SU}(z;\tau,\bar\tau) = 2 (z^{-1}\! +1+z) \cG_{SU}(z;\tau,\bar\tau)+2(z-1)(2z+1) \partial_z \cG_{SU}(z;\tau,\bar\tau)+(z-1)^2 z \partial_z^2 \cG_{SU}(z;\tau,\bar\tau)\, .
 \label{lapnew} 
 \eea

\subsection{Derivation from the hermitian matrix model}
\label{sec:matrix0}

 We will now prove that the generating function is indeed given by the conjectured form \eqref{eq:gresult}, and that the integrated correlator,  $\cC_{G_N}(\tau,\bar\tau)$, satisfies the Laplace difference equation \eqref{lapdiffSUN}, using properties of correlation functions in the $N\!\times\! N$ hermitian matrix model. Our procedure will be based on the analysis of the perturbative contribution to the integrated correlator $\cC_{G_N}(\tau,\bar\tau)$. It was argued in \cite{Dorigoni:2021guq, Dorigoni:2022zcr} (see also \cite{Collier:2022emf}) that once the correlator is known to have the form in \eqref{gsun}, the functions $B^i_{G_N}(t)$ are completely determined by the perturbative contributions to $\cC_{G_N}(\tau,\bar\tau)$. Given complete knowledge of $\cG_{SU(2)}(\tau,\bar\tau)$  \cite{Dorigoni:2021guq}, this will uniquely determine the $SL(2,\Z)$ invariant form for $\cG_{G_N}(\tau,\bar\tau)$ .
 Although we will continue to present the $SU(N)$ case in this section, the results extend straightforwardly to a general classical gauge group, $G_N$. 
 
   As shown in \cite{Dorigoni:2022zcr}, the perturbative terms lead to the relation
\ie \label{eq:BSUN2}
B_{SU(N)}(t) &= - t \int_0^{\infty}  e^{-x t}  x^{3\over 2} \partial_x \left[ x^{3\over 2} \partial_x  I_{SU(N)}(x) \right] dx \, ,
\fe
where the function $I_{SU(N)}(x)$ is defined in  \eqref{pertSUn2}  and determines the perturbative part  of $\cC^{pert}_{SU(N)}(\tau,\bar\tau)$ via \eqref{pertSUn}.
The generating function $\cB_{SU}(z; t)$ can then be expressed as
\ie \label{eq:gBN}
\cB_{SU}(z; t)  =- t \int_0^{\infty} e^{-x t}  x^{3\over 2} \partial_x \left[ x^{3\over 2} \partial_x  I_{SU}(z; x) \right]dx  \, ,
\fe
where $I_{SU}(z;x):= \sum_{N=1}^\infty I_{SU(N)}(x)\, z^N$. As explained in appendix \ref{app:matrix1}, $I_{SU(N)}(x)$ can be obtained from a specific combination of one-point and two-point correlation functions in the $N\!\times\! N$  hermitian matrix model \cite{Chester:2019pvm}, as reviewed in  \eqref{eq:INeN}. Using this relation and the expression for the generating functions of  the hermitian matrix model correlators given in \eqref{eq:ge1} and \eqref{eq:ge2}, we have
\ie
I_{SU}(z; x) =&  {z\over (1-z)^2}  \left[ \left(1- e^{-x}  \right)  - \oint\!\! \! \oint { z \over u_1 u_2 (u_1 u_2 -z)} \exp \left( {x  \over 2} {u_1 u_2 -1 \over (1-u_1)(1-u_2)} \right) {du_1 \over 2\pi i} {du_2 \over 2 \pi i} \right] \cr
&- {1\over x^2}  \oint\!\! \! \oint { z (u_1 u_2+z) \over  (u_1 u_2 -z)^3} \exp \left( {x  \over 2} {u_1 u_2 -1 \over (1-u_1)(1-u_2)} \right)  {du_1 \over 2\pi i} {du_2 \over 2 \pi i}\, ,
\label{eq:matrixrels}
\fe
where   the first line is the contribution arising from the matrix model  two-point  function  and the second is from the square of the one-point function. Upon performing the $u_2$ contour integral around $u_2=0$ and $u_2=z/u_1$ the result  is 
\ie
I_{SU}(z; x) =&  {z\over (1-z)^2}  \left[ 1 -   \oint {1\over u_1}   \exp \left( -x { u_1  (z-1) \over (1-u_1)(z-u_1)} \right) {du_1\over 2\pi i} \right] \cr
&-  \oint {z (u_1^2 - u_1 x z- z^2) \over x(u_1-z)^4} \exp \left( -x { u_1  (z-1) \over (1-u_1)(z-u_1)} \right)   {du_1 \over 2\pi i} \, .
\fe
Substituting the above expression for $I_{SU}(z; x)$ in  \eqref{eq:gBN} and performing the $x$ integral leads to an expression that has a pole at $t (u_1-1) (u_1-z)+ u_1(1-z)=0$.  The $u_1$ contour integral picks up the residue at this pole, and gives the  previously conjectured expression \eqref{eq:gresult}.
 
The above argument provides a proof that $\cB_{SU}(z; t)$ obeys the differential equation \eqref{eq:lapdiffgsu}.
This proof was based on an analysis of the perturbative terms in the localised integrated correlator (these are the terms that arise from the one-loop determinant in Pestun's analysis  \cite{Pestun:2007rz}).   This  automatically ensures that the perturbative part of  $\cC_{SU(N)}(\tau,\bar\tau)$ satisfies the Laplace difference equation \eqref{lapdiffSUN}  with 
the Laplace operator replaced by  $\tau_2^2 \partial^2_{\tau_2}$, since perturbation theory is independent of  $\tau_1$.  

Invariance under $SL(2, \mathbb{Z})$ is restored in a unique manner by simply extending the differential operator $\tau_2^2 \partial^2_{\tau_2}$ to the Casimir operator $\Delta_{\tau} = \tau_2^2(\partial_{\tau_1}^2+\partial_{\tau_2}^2)= 4\tau_2^2\partial_\tau\partial_{\bar{\tau}}$, and the perturbative Laplace difference equation leads to \eqref{lapdiffSUN}. Importantly, in \cite{Dorigoni:2021guq} it was explicitly verified that the initial term $\C_{SU(2)}(\tau, \bar \tau)$ is indeed given by \eqref{gsun}. Therefore, with this initial condition and the recursion relation \eqref{lapdiffSUN}, the $SL(2,\Z)$ invariant expression  \eqref{gsun}  for $\C_{SU(N)}(\tau, \bar \tau)$ follows. 

\subsection{Fourier mode decomposition of the generating function}
\label{sec:fourier}

We will now comment on some properties of the generating function, in particular its Fourier expansion with respect to $\tau_1$.  This   is obtained by performing a Poisson resummation that transforms the sum over $m$  in \eqref{eq:cGSU} into a sum over $\hat m$, \footnote{Recall that Poisson resummation transforms  $\sum_{m\in \Z}f(m)$ into $\sum_{\hat m\in \Z} \hat{f}(\hat m)$ where $\hat{f}$ denotes the Fourier transform of $f$ and it is given by $\hat{f}(\hat{m}) = \int_\mathbb{R} e^{-2\pi i m \hat m}   f(m)dm$.}
giving 
\ie
   \cG_{SU}(z;\tau,\bar\tau)  = \sum_{k\in\Z}e^{2\pi i k\tau_1} \cG^{(k)}_{SU}(z;\tau_2) =  \sum_{(\hat{m},n)\in\Z^2} e^{2\pi i \hat{m}n \tau_1} \int_0^\infty e^{-\pi t n^2 \tau_2 -\frac{\pi \hat{m}^2 \tau_2}{t} } \frac{\sqrt{\tau_2} \cB_{SU}(z;t)}{\sqrt{t}} dt\,.
   \label{zeromodetwo}
\fe
  In order to analyse the large-$N$ expansion it will prove useful to transform \eqref{zeromodetwo} from an integral over $t$ in the range $(0,\infty)$ to the range $(1,\infty)$.  
    
For this purpose we  split the $t$ integral into the domains $t\in(0,1)$ and $t\in(1,\infty)$.
The change of variables $t\to t^{-1}$ maps the $(0,1)$ interval into $(1,\infty)$.   We may then  rewrite this integral using the inversion property $t^{-1}\cB_{SU}(z;t^{-1} )  = \cB_{SU}(z; t )$, together with the interchange $(\hat{m},n)\to (n,\hat{m})$. 
This results in the following Fourier series,
 \begin{align}
  \cG_{SU}(z;\tau,\bar\tau)  & = 2 \sum_{(\hat{m},n)\in\Z^2} e^{2\pi i \hat{m}n  \tau_1}\int_1^\infty  e^{-\pi t n^2 \tau_2 -\frac{\pi \hat{m}^2 \tau_2}{t} } \frac{\sqrt{\tau_2} \cB_{SU}(z;t)}{\sqrt{t}} dt \nn\\
  & = 2 \sum_{(m,n)\in\Z^2} \int_1^\infty  e^{- t \,Y_{mn}(\tau,\bar\tau) }  \cB_{SU}(z;t) dt\,,
    \label{eq:fourierm}
 \end{align}
 where the second line follows from a Poisson resummation back to the original variables $(\hat{m},n)\to (m,n)$.
   
For future reference, we note that the zero Fourier mode of $\cC_{SU}(\tau,\bar\tau)$ is given by the sum of two kinds of terms.   The first kind consists of a sum of terms in the second line of  \eqref{eq:fourierm}  with $m=\ell \in \Z$, $n=0$.  The second kind consists of a sum of  terms with $\hat m=0,n=\ell\in \Z$ in the first line of  \eqref{eq:fourierm},  which is equivalent to a sum over all $m\in \Z$ and $n=\ell\in \Z$ with  $\ell \ne 0$.
The resulting zero mode can be expressed as   
 \begin{align}  \label{eq:0mode} 
  \cG^{(0)}_{SU}(z;\tau_2) &= 2 \int_1^\infty \cB_{SU}(z;t) dt+4 \sum_{\ell=1}^\infty \int_1^\infty \Big[  e^{-\frac{\pi t \ell^2 }{\tau_2}  }+e^{-\pi t \ell^2 \tau_2 } \frac{\sqrt{\tau_2}}{\sqrt{t}}\Big] \cB_{SU}(z;t) dt \,,
 \end{align}
 where we  have used the property
 \begin{equation}
 \int_1^\infty \frac{\sqrt{\tau_2}}{\sqrt{t}} \cB_{SU}(z;t)dt  = \int_0^1 \frac{\sqrt{\tau_2}}{\sqrt{t}} \cB_{SU}(z;t)dt  = 0\,.
 \end{equation} 
 We also note
 \begin{align}
  \int_1^\infty \cB_{SU}(z;t) dt &=\frac{z\big( 1+z-(1-z)\sqrt{1-z}\big)}{8 (1-z)^3}= \sum_{N=1}^\infty \Big[ \frac{N^2}{8}-\frac{\Gamma(N+1/2)}{4\sqrt{\pi}\Gamma(N)}\Big]z^N := \sum_{N=1}^\infty c_{1}(N) z^N \,,\label{eq:const1infty}\\
   \int_0^1 \cB_{SU}(z;t) dt&=-\frac{z \big(1-z- (1-z)\sqrt{1-z} \big)}{8(1-z)^3} =\sum_{N=1}^\infty \Big[- \frac{N}{8}+\frac{\Gamma(N+1/2)}{4\sqrt{\pi}\Gamma(N)}\Big]z^N:=\sum_{N=1}^\infty c_{2}(N) z^N\,.\label{eq:const01}
 \end{align}
 We will return to these expressions when we consider the large-$N$ expansion in the next two sections.

 \section{Large-$N$ expansion at fixed $\tau$}
 \label{sec:largeN}
 
The large-$N$ expansion of the integrated four-point  correlator has a close relation to the $\alpha'$-expansion of the integrated  four-graviton amplitude in $AdS_5\times S^5$. These properties were elucidated  in  \cite{Binder:2019jwn,Chester:2019jas, Chester:2020dja}  where the $1/N$ expansion was considered in both the 't Hooft limit (in  which $\lambda= g_{_{Y\!M}}^2 N$ is fixed) and in the limit in which $g_{_{Y\!M }}$ is fixed.  In this section we will see that  the large-$N$ expansion of $\cC_{SU(N)}(\tau,\bar\tau)$  is streamlined when expressed in terms of the integral representation given by the contour integral
 \begin{equation}
 \label{eq:sunintcorr}
\cC_{SU(N)}(\tau,\bar\tau)= 2\sum_{(m,n)\in \Z^2} \int_1^\infty e^{-t  \,Y_{mn}(\tau, \bar \tau) } \Big[ \oint_C \frac{\cB_{SU}(z;t)}{z^{N+1}}  \frac{dz}{2\pi i } \Big] dt \,,
 \end{equation}
 as follows from \eqref{eq:geninv}  and the second line of  \eqref{eq:fourierm}.    Furthermore, we will see  that this provides a natural procedure for determining the non-perturbative terms that complete the  $1/N$ expansion.

\begin{figure}[t!]
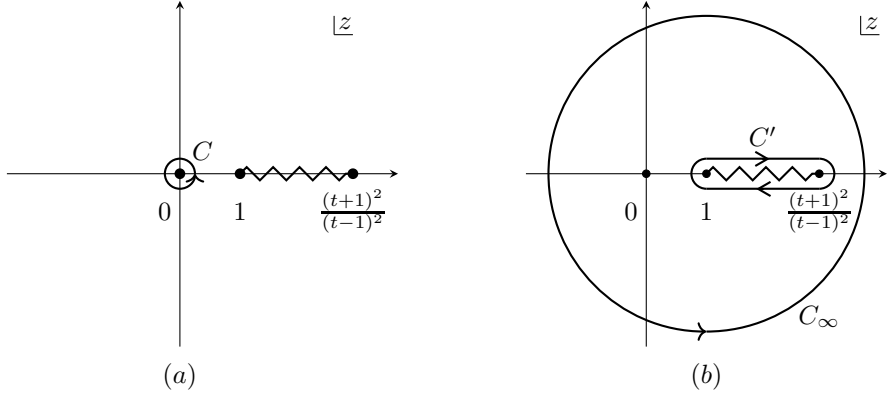

\begin{center}
\tikzpicture[scale=1.0]
\scope [very thick, every node/.style={sloped,allow upside down},xshift=-03.5cm,yshift=0cm]
%
%

\draw[thin] (-2.3,0) -- (2.3,0) ;
\draw[-stealth,thin] (0,-2.3) -- (0,2.3) ;]
\draw [fill=black] (-0.0,0.0) circle [radius=.05];
\draw[thick, -> ]     (.2,0) arc (0:360:0.2cm);
\draw (2.2, 2.0) node{$z$};
\draw[thin] (2.05,1.85) -- (2.3,1.85) ;
\draw[thin] (2.05,1.85) -- (2.05,2.10) ;

\draw [fill=black] (2.3,0.0) circle [radius=.05];
\draw [fill=black] (0.8,0.0) circle [radius=.05];
\draw[snake,thick] (0.8,0) -- (2.3,0) ;  
\draw (-0.2, -0.5) node{$0$};
\draw (0.30, 0.30) node{$C$};
\draw (0.8, -0.5) node{$1$};
\draw (2.3, -0.5) node{$  \frac{(t+1)^2}{(t-1)^2}$};
\draw [-stealth,thin](2.3,0) -- (2.9,0.0) ;
\draw (0, -2.7) node{$(a)$};
\endscope 
\scope[xshift=3.5cm,yshift=0cm]
\draw[thick, ->] (0,-2.1) arc (-90:270:2.1cm);
\draw[ thick, ] (0.0,-0.2) arc (270:90:0.2cm);
\draw[ thick ] (1.5,0.2) arc (90:-90:0.2cm);
\draw [thin](-2.4,0) -- (0,0) ;
\draw [-stealth,thin](1.5,0) -- (2.4,0.0) ;
\draw[-stealth,thin] (-0.8,-2.3) -- (-0.8,2.3) ;]
\draw [fill=black] (1.5,0.0) circle [radius=.05];
\draw [fill=black] (-0.0,0.0) circle [radius=.05];
\draw[snake,thick] (0,0) -- (1.5,0) ;  
\draw[thick] (0,0.2) --   node {\midarrow}  (1.5,0.2)  ; 
\draw[thick] (1.5,-0.2)--   node {\backmidarrow}  (0,-0.2)   ; 
\draw (-1, -0.5) node{$0$};
\draw (0, -0.5) node{$1$};
\draw (1.5, -0.5) node{$  \frac{(t+1)^2}{(t-1)^2}$};
\draw (0.75, 0.5) node{$C'$};
\draw (1.5, -1.9) node{$C_\infty$};
\draw [fill=black] (-0.8,0.0) circle [radius=.05];
\draw (0, -2.7) node{$(b)$};
\draw (2.2, 2.0) node{$z$};
\draw[thin] (2.05,1.85) -- (2.3,1.85) ;
\draw[thin] (2.05,1.85) -- (2.05,2.10) ;
\endscope
\endtikzpicture
\end{center} 
\caption{(a) The contour $C$ encircling the pole at $z=0$. 
(b) The distorted contour $C'$ encircles the cut, together with the contour at infinity, $C_\infty$, which gives a vanishing contribution. 
}
\label{fig:contour}
\end{figure}

We proceed by splitting  the sum in \eqref{eq:sunintcorr} into the $(m,n)=(0,0)$ component and the rest
 \begin{equation}\label{eq:Lattice1}
\cC_{SU(N)}(\tau,\bar\tau)=2 c_1(N)+ 2\!\!\sum_{(m,n)\neq (0,0)} \int_1^\infty e^{-t \, Y_{mn}(\tau, \bar \tau)  } \Big[ \oint_C \frac{\cB_{SU}(z;t)}{z^{N+1}}  \frac{dz}{2\pi i } \Big] dt \,,
 \end{equation}
 where the constant term $c_1(N)$ is given in \eqref{eq:const1infty}.
 The integration contour $C$, shown in figure \ref{fig:contour}(a), can be distorted into the sum of the contour at infinity, $C_\infty$, together with the contour $C'$ surrounding the branch cut, as shown in figure \ref{fig:contour}(b).
The contour at infinity does not contribute since $\cB_{SU}(z;t) = O(1/z^2)$ as $|z|\to \infty$.
The resulting integral is given by 
  \begin{equation}
\cC_{SU(N)}(\tau,\bar\tau)=2 c_1(N)+ 2\!\!\sum_{(m,n)\neq (0,0)} \int_1^\infty e^{-t \, Y_{mn}(\tau, \bar \tau) } \Big[\oint_{C'} \frac{\cB_{SU}(z;t)}{z^{N+1}}  \frac{dz}{2\pi i } \Big] dt \,,\label{eq:GNint}
 \end{equation}
 where the new contour of integration $C'$ is a clockwise contour surrounding the branch-cut located on the interval $z\in [1, \frac{(t+1)^2}{(t-1)^2}]$.
The discontinuity across the branch-cut can easily  be computed
\begin{align}
\label{gsures}
\mbox{Disc} \cB_{SU}(z;t) &= \lim_{\epsilon\to 0^+}\Big[ \cB_{SU}(z+i\epsilon;t) - \cB_{SU}(z-i\epsilon;t) \Big]\\
&\nn = - 2i   \frac{3 t z^2 \left[(t-3) (t+1)^2 (3 t-1)-(t-1)^2
   (t+3) (3 t+1) z\right]}{2 (z-1)^{\frac{3}{2}}
   \left[(t+1)^2-(t-1)^2 z\right]^{\frac{7}{2}}}\,,\qquad z\in \left[1, \frac{(t+1)^2}{(t-1)^2} \right]\,.
\end{align}
Although we would like to write
\begin{equation}
B_{SU(N)}(t)= \oint_{C'} \frac{\cB_{SU}(z;t)}{z^{N+1}}  \frac{dz}{2\pi i }  =  \int_1^{\frac{(t+1)^2}{(t-1)^2}} \frac{\mbox{Disc} \cB_{SU}(z;t)}{z^{N+1}}  \frac{dz}{2\pi i }\,,
\end{equation} 
the discontinuity is not quite an integrable function due to the end-point singularities $(z-1)^{-\frac{3}{2}}$ and $(z_1-z)^{-\frac{7}{2}}$ with
\begin{equation}
z_1:=\frac{(t+1)^2}{(t-1)^2}\,.\label{eq:z1}
\end{equation}
However if we regularise the integral by replacing the singular end-point factors by $(z-1)^{-\alpha}$ and $(z_1-z)^{-\beta}$ and then take the limit $\alpha\to3/2$ and $\beta\to 7/2$ after integration, the result is 
\begin{align}
&B_{SU(N)}(t)=\! \lim_{\alpha\to \frac{3}{2}\,,\,\beta\to \frac{7}{2}} \int_1^{z_1} (-2i)\frac{3 t z^2 \left[(t-3) (t+1)^2 (3 t-1)-(t-1)^2
   (t+3) (3 t+1) z\right]}{2 (z-1)^{\alpha}
   \left[(t+1)^2-(t-1)^2 z\right]^{\beta} z^{N+1}}  \frac{dz}{2\pi i } \\
 & = \frac{N(N^2-1)t }{4(t-1)^7} \Big[(t-1)^2 (t+3)(3t+1){}_2F_1\Big(\frac{7}{2},N+2;4\vert 1-z_1\Big)\!-\!2(N+2)t(t^2+1){}_2F_1\Big(\frac{7}{2},N+3;5\vert 1 -z_1\Big) \Big].  \nn
\end{align}
This expression is perfectly regular and it is straightforward to check that it reproduces the correct answer since it  is identical to \eqref{eq:BSUN}  for any value of $N\in \mathbb{N}$.

 \subsection{Saddle-point analysis}
 \label{sec:sad-pt}
  
In order to analyse the behaviour of $\cC_{SU(N)}(\tau,\bar\tau)$ at large $N$ we will find it convenient to consider the distinct contributions of different  regions of the $z$ integration.  We will see that the region near the endpoint  $z\sim 1$ produces the series of terms that are perturbative in $1/N$, while the region near the endpoint $z\sim z_1$ produces terms that are exponentially suppressed in $N$.

To see this it is convenient to perform the change of variables $z = e^{\mu}$ and consider 
\begin{align}\label{eq:ztomu}
 B_{SU(N)}(t) = \,& \int_1^{z_1}  \frac{\mbox{Disc} \cB_{SU}(z;t)}{z^{N+1}}  \frac{dz}{2\pi i }=\int_0^{\log z_1} e^{- N \mu} \,\frac{\mbox{Disc} \cB_{SU}(e^{\mu};t)}{2\pi i } d\mu\,.
 \end{align}
 We will separate this integral into the sum of two pieces by writing $B_{SU(N)}(t)= B^{P}_{SU(N)} (t)+ B^{N\!P}_{SU(N)} (t)$ where
 \begin{equation}
 B_{SU(N)}^{P} (t):= \int_0^{\infty\pm i\epsilon}  e^{- N \mu}\, \frac{\mbox{Disc} \cB_{SU}(e^{\mu};t)}{2\pi i }d\mu \, , \label{eq:I12adef}
 \end{equation}
and
 \begin{equation}
\,\,\, B_{SU(N)}^{N\!P} (t) := -   \int_{\log z_1}^{\infty \pm i\epsilon}   e^{-N \mu} \,\frac{\mbox{Disc} \cB_{SU}(e^{\mu};t)}{2\pi i }d\mu  \,,
\label{eq:I12def}
 \end{equation}
 where the superscripts ${P}$ and ${N\!P}$ indicate the pieces that contain terms perturbative and non-perturbative in $1/N$ in the large-$N$ regime.
 
  Note that the discontinuity $\mbox{Disc} \cB_{SU}(z;t)$ in \eqref{gsures} itself has a discontinuity along the interval $z\in[z_1,\infty)$ with $z_1$ given in \eqref{eq:z1}. For this reason, when we extend the upper limit of the domain of integration from $z\in[0,z_1]\to z\in [0,\infty\pm i \epsilon)$ in  \eqref{eq:I12adef}, we have to specify on which side of the branch cut we are integrating. Equation \eqref{eq:I12adef} can be thought of as a lateral Borel resummation with respect to the complexified parameter $N$, where the integrand $\mbox{Disc} \cB_{SU}(e^\mu;t)/(2\pi i)$ plays the r\^ole of Borel transform for the large-$N$ perturbative expansion (see e.g. \cite{Dorigoni:2014hea} for a recent introduction to resurgence). 

 The $\pm i \epsilon$ deformation in the upper limit of  \eqref{eq:I12adef} will become irrelevant when we only consider the formal, asymptotic large-$N$ expansion of \eqref{eq:I12adef}. As expected from resurgence theory, both lateral resummations defined in \eqref{eq:I12adef} will give rise to the same formal asymptotic power series in $N$. However, precisely due to the branch-cut singularity of the integrand $\mbox{Disc} \cB_{SU}(e^\mu;t)/(2\pi i)$, there is an ``ambiguity'' in resummation of the perturbative expansion, reflected by the $\pm i \epsilon$ in \eqref{eq:I12adef}. This ambiguity in resummation has to be compensated by the non-perturbative corrections which are fully encoded in \eqref{eq:I12def}, which in resurgence language can be related to the Stokes automorphism. 
 
 From a resurgence point of view, when we correlate the resummation of the perturbative expansion, \eqref{eq:I12adef}, with the corresponding non-perturbative corrections, \eqref{eq:I12def}, we obtain the unambiguous and exact result \eqref{eq:ztomu},
 usually called \textit{median resummation}.
As will become clear shortly, this is a manifestation of the similar ambiguity and median resummation in the 't Hooft coupling resurgent expansion described in \cite{Dorigoni:2021guq}.
 
 Let us first focus on re-deriving the large-$N$ perturbative expansion, previously derived in \cite{Dorigoni:2021guq,Dorigoni:2021bvj}.
 Rescaling  $\mu\to \mu/N$ and then expanding ${\rm Disc}\cB_{SU}(e^{\frac{\mu}{N}};t)$ around the point $\mu/N=0$ (i.e.~near $z= 1$) leads to  
 \ie \label{eq:largeNdisc} 
 &\mbox{Disc} \cB_{SU}(e^{\frac{\mu}{N}};t)  = i \frac{3 N^{\frac{3}{2}}(t^{-\frac{3}{2}}+t^{\frac{1}{2}})}{16 \mu^{\frac{3}{2}}} +i \frac{15 N^{\frac{1}{2}}(t^{-\frac{5}{2}}+t^{\frac{3}{2}})}{64 \mu^{\frac{1}{2}}}+i \frac{3\mu^{\frac{1}{2}}[105(t^{-\frac{7}{2}}+t^{\frac{5}{2}})-13 (t^{-\frac{3}{2}}+t^{\frac{1}{2}})]}{2048 N^{\frac{1}{2}}}+ O(N^{-\frac{3}{2}})\, .
 \fe
 We now substitute this expansion into the integral  for $B_{SU(N)}^P(t)$  \eqref{eq:I12adef}   taking care to make the modification
$\mu^{-\frac{3}{2}}\to \mu^{-\alpha}$,  as explained earlier.   Setting $\alpha=\frac{3}{2}$ after performing the $\mu$ integral we obtain 
 \begin{equation}\label{eq:B1pert}           
  B_{SU(N)}^P(t) =  -N^{\frac{1}{2}}\frac{3(t^{-\frac{3}{2}}+t^{\frac{1}{2}})}{16\sqrt{\pi}} + \frac{15(t^{-\frac{5}{2}}+t^{\frac{3}{2}})}{128 \sqrt{\pi} N^{\frac{1}{2}}}  +\frac{[315(t^{-\frac{7}{2}}+t^{\frac{5}{2}}) - 39 (t^{-\frac{3}{2}}+t^{\frac{1}{2}})]}{8192 \sqrt{\pi} N^{\frac{3}{2}}}+O(N^{-\frac{5}{2}})\,,
 \end{equation}
 which agrees precisely  with equation (5.56) in \cite{Dorigoni:2021guq} (allowing for a factor of 2 change in our normalisation conventions).

Upon substituting this asymptotic expansion into \eqref{eq:Lattice1} we see that the constant term $c_1(N)$, defined in \eqref{eq:const1infty}, combines with a Dirichlet regularisation for the lattice-sum in the zero-mode sector, thereby   reproducing the correct constant $N^2/4$. What is left leads to an infinite series of terms that contribute to the large-$N$ perturbative terms of $\cC_{SU(N)}(\tau,\bar\tau)$. These remaining non-constant terms are power behaved in $1/N$ and with coefficients given by finite  rational sums of non-holomorphic Eisenstein series of half-integral index, reproducing the results of \cite{Dorigoni:2021guq,Dorigoni:2021bvj}:\footnote{Note that thanks to the results of section \ref{sec:fourier}, the $t$ integral now runs over $(1,\infty)$ rather than $(0,\infty)$ so that both the $t^{s}$ and $t^{1-s}$ terms in \eqref{eq:B1pert} are integrable when multiplied by $e^{-t Y_{mn}(\tau,\bar\tau)}$ with $(m,n)\neq(0,0)$.} 
\ie \label{eq:pSU}
\cC^{P}_{SU(N)} (\tau, \bar \tau)  &=   2c_1(N)+2\sum_{(m,n)\neq(0,0)} \int_1^\infty e^{-t Y_{mn}(\tau,\bar\tau)} B_{SU(N)}^{P}(t) dt\cr
 &  = {N^2 \over 4} +  \sum_{r=0}^{\infty} N^{{1\over 2}-r}  \sum^{\lfloor r/2 \rfloor}_{m=0} b_{r, m}E(\threeh {+} \delta_r {+} 2m; \tau, \bar \tau)  \, ,
 \fe
 where $\delta_r=0$ for even $r$ and $\delta_r=1$ for odd $r$ and the coefficients $b_{r,m}$ can be found in those references, or can be obtained from $B_{SU(N)}^P(t)$ as expanded in \eqref{eq:B1pert}.
 
As previously mentioned, the large-$N$ expansion of $ B_{SU(N)}^P(t)$ produces a purely perturbative yet formal, asymptotic power series at large $N$ and it is insensitive to the $\pm i \epsilon$ deformation of the contour of integration  \eqref{eq:I12adef}.
The ambiguity in resumming \eqref{eq:B1pert} to \eqref{eq:I12adef}, or equivalently in resumming \eqref{eq:pSU}, is compensated by the change in non-perturbative corrections exponentially suppressed in $N$ at large $N$ and fully captured by $B_{SU(N)}^{N\!P}(t)$ in \eqref{eq:I12def}. 
 This may be exhibited as follows,
 \begin{align}
 \label{eq:bn2int}
B_{SU(N)}^{N\!P}(t) &=   \int_{z_1}^{\infty\pm i \epsilon}  \frac{\mbox{Disc} \cB_{SU}(z;t)}{z^{N+1}}  \frac{dz}{2\pi i }\\
&\nn =   (\pm i)  \lim_{\beta\to \frac{7}{2}}\int_{z_1}^\infty (-2i)\frac{3 t z^2 \left[(t-3) (t+1)^2 (3 t-1)-(t-1)^2
   (t+3) (3 t+1) z\right]}{2 (z-1)^{\frac{3}{2}}
   \left[-(t+1)^2+(t-1)^2 z\right]^{\beta} z^{N+1}} \frac{dz}{2\pi i }\,.
 \end{align}

Substituting this expression into \eqref{eq:GNint}  leads to a $t$ integral that is dominated by a saddle point when $N \gg Y_{mn}(\tau,\bar\tau)$. This integral has the form
 \begin{equation}
\int_1^\infty h(N,t)  e^{-t \, Y_{mn}(\tau,\bar\tau)}z_1^{-N} dt = \int_1^\infty h(N,t) \exp\Big[-t \, Y_{mn}(\tau,\bar\tau) - 2N \log\Big( \frac{t+1}{t-1} \Big) \Bigr] dt \,,
\label{eq:expact}
 \end{equation}
 where we have expressed $B_{SU(N)}^{N\!P}(t)$ as $h(N,t) z_1^{-N}$ (note $z_1 =\frac{(t+1)^2}{(t-1)^2}$ from \eqref{eq:z1}), and $h(N,t)$ contains only power-behaved terms at large $N$.  The integrand has two saddle points  located at
 \begin{equation}\label{eq:saddles}
 t^\star_{1} =  \frac{\sqrt{4 N+ Y_{mn}(\tau,\bar\tau)}}{\sqrt{ Y_{mn}(\tau,\bar\tau)}}\, , \qquad \qquad  t^\star_{2} = - \frac{\sqrt{4 N+ Y_{mn}(\tau,\bar\tau)}}{\sqrt{ Y_{mn}(\tau,\bar\tau)}}\,.
 \end{equation}
 A simple thimble analysis shows that only the saddle $t^\star_1$ is connected with the contour of integration of interest.\footnote{This is a consequence of the fact that the $t$ integral was defined to span the domain $(1,\infty)$ as discussed in section \ref{sec:fourier}.}  The ``on-shell'' expression for such a saddle is given by
  \begin{equation} 
 \exp\Big[-t^\star_1  Y_{mn}(\tau,\bar\tau) - 2N \log\Big( \frac{t^\star_1+1}{t^\star_1-1} \Big) \Big] = \exp\Big[-N A\Big(\sqrt{ \frac{ Y_{mn}(\tau,\bar\tau)}{4N}}\Big)\Big]\,
 ,\label{eq:modterms}
 \end{equation}
 where the function $N A(x)$ is the saddle-point action and $A(x)$ is given by
 \begin{equation}
 A(x) := 4\Big( x\sqrt{x^2+1} +\mbox{arcsinh}(x)\Big)\,.
 \label{eq:Action}
  \end{equation}
This expression is identical to that recently found in a resurgence analysis of the Fourier zero mode of the integrated correlator  in \cite{Hatsuda:2022enx}.   As pointed out in that paper, the function  $A(x)$ coincides with the $D3$-brane action  discussed in \cite{Drukker:2005kx} in the evaluation of Wilson loops in large representations of the $SU(N)$ gauge group. We will return to a discussion of this connection shortly. For now, we note the behaviour of $A(x)$ in the small $x$ is given by,
  \bea
A(x) = 8x+\frac{4}{3}x^3-\frac{1}{5} x^5 + \ldots \, .
  \label{eq:Alargex}
  \eea
  Since the lattice sum over $(m,n)$ in \eqref{eq:GNint} is convergent, in the large-$N$ limit with $\tau$ fixed we may expand the summand first. In particular, to leading order we have
 \begin{equation}
 \exp\Big[-N A\Big(\sqrt{ \frac{ Y_{mn}(\tau,\bar\tau)}{4N}}\Big)\Big] \stackrel{N\to\infty}{\sim}\exp(-4 \sqrt{N Y_{mn}(\tau,\bar\tau)}) =  \exp\Big(-4 \sqrt{\frac{N \pi}{\tau_2} } |m+n\tau|\Big)\,.\label{eq:modeLarge}
 \end{equation}

 In order to understand the behaviour more generally we need to determine fluctuations   of the exponent (or the ``action'')  in \eqref{eq:expact}
   \begin{equation}
 S(t) := t\,  Y_{mn}(\tau,\bar\tau)+2 N\log\Big(\frac{t+1}{t-1}\Big)\,,\label{eq:Sos}
 \end{equation}
  around the saddle point value $t=t_1^\star$.  Denoting the  fluctuation of $t$   by $ t= t_1^\star+ N^{\frac{1}{4}} \delta\,$  we have 
 \begin{equation}
 S(t_1^\star+ N^{\frac{1}{4}} \delta)   = S(t_1^\star)+k^2 \delta^2+O( N^{-\frac{1}{4}}\delta^3)\,,
 \label{eq:deltafluc}
 \end{equation}
 where $ t_1^\star$ is given in \eqref{eq:saddles}, the on-shell action $S(t_1^\star)$ appears in \eqref{eq:modterms}, and
 \begin{equation}
k^2 = \frac{ \big(Y_{mn}(\tau,\bar\tau)\big)^{\frac{3}{2}}}{2}(1+O(N^{-1}))\, . 
 \end{equation}

Upon  expanding the  exponential of the action  in powers of $\delta$ and performing gaussian integrals over $\delta$, we obtain the exponentially suppressed terms in the large-$N$ limit.  In the $1/N$ expansion, these terms are given by a sum over new non-holomorphic modular invariant functions $D_N(s; \tau, \bar \tau)$, 
\ie \label{eq:npSU}
\cC^{N\!P}_{SU(N)} (\tau, \bar \tau)  &=   2\sum_{(m,n)\neq(0,0)} \int_1^\infty e^{-t Y_{mn}(\tau,\bar\tau)} B_{SU(N)}^{N\!P}(t) dt\cr
 &=   \pm i    \sum_{r=0}^{\infty} N^{2-\frac{r}{2}} \sum_{m=0}^{r} d_{r,m} D_N\big(2m-\frac{3r}{2} ; \tau, \bar \tau\big) \, ,
\fe
where $D_N(s; \tau, \bar \tau)$ takes the following form
\ie
D_N(s; \tau, \bar \tau)  & :=  \sum_{(m,n)\neq (0,0)} \exp\Big(- 4 \sqrt{N  Y_{mn}(\tau, \bar \tau)}\Big)  \left( Y_{mn} (\tau, \bar \tau) \right)^{-s}    \\
&=  \sum_{\ell = 1}^\infty \sum_{{\rm gcd} (p,q)=1 } \exp\Big(- 4 \sqrt{N \pi} \ell \frac{|p+q\tau|}{\sqrt{\tau_2}} \Big) \frac{1}{\pi^s} \frac{\tau_2^s }{\ell^{2s} |p+q\tau|^{2s}}\,.
\label{eq:Ddef}
 \fe
Some basic properties of $D_N(s; \tau, \bar \tau)$ are discussed in the appendix \ref{app:newmod}.
The explicit form of the first few coefficients $d_{r,m}$ in \eqref{eq:npSU}   are
\ie\label{eq:coeffdrm}
d_{0, 0} &= -2\, ,  \cr
 d_{1, 0} &={1\over 3}\, ,  \quad \,\,\, \,\,\, \,\,d_{1, 1} =-{9\over 4}\, ,\cr
 d_{2, 0} &= -{1\over 36}\, ,\quad  d_{2, 1} ={3\over 8}\, , \quad  \quad\,\,\,\,\, d_{2, 2} =-{117\over 64}\, , \cr
d_{3,0} &= {1\over 648} \, , \quad\,\, d_{3,1} =-{7\over 160} \, , \quad  d_{3,2} =-{77\over 128}\, ,  \quad  d_{3, 3} =-{489 \over 512} \, .
\fe
In fact, once the leading coefficients $d_{r, r}$ are specified all other coefficients are determined by the Laplace difference equation \eqref{lapdiffSUN}.\footnote{The action of Laplace operator on the modular function $D_N(s; \tau, \bar \tau)$ is given in \eqref{eq:LapD}. }  This is very similar to the pattern of coefficients of the perturbative series of $N^{\frac{1}{2}-r}$ terms in  \eqref{eq:pSU} for which all $b_{r,m}$ are determined by the  Laplace difference equation, except $b_{r, \lfloor r/2 \rfloor}$, which has to be specified \cite{Dorigoni:2021guq}.   In other words the coefficients of the  terms in the  second line of \eqref{eq:genN}  are all determined once we input the coefficients with value of $m=0$. Furthermore, as was noted in \cite{Dorigoni:2021guq},   the perturbative coefficients $b_{r, \lfloor r/2 \rfloor}$ are  determined by the leading $N^2$ term in the 't Hooft limit. The same is true for the coefficients $d_{r, r}$ that are also determined by the $N^2$ term in the 't Hooft limit.  Such a  contribution was determined in \cite{Dorigoni:2021guq} using resurgence methods\footnote{The 't Hooft limit of \eqref{eq:npSU} is discussed in details in section \ref{sec:tHooft} and appendix \ref{app:zeromode}.}, and it was shown (equation (5.39) of \cite{Dorigoni:2021guq}) that $d_{r, r}$
is given by 
\ie
d_{r,r} = -\frac{ a_r}{4^{r+1}} \, ,
\fe 
where $a_{r}$ is determined by the following recursion relation, 
\ie
r (r-4) (r+2) (2 r^2 + 2 r -9) a_{r} + 
 3 (2 r^4 - 17 r^2  + 9 r + 39) a_{r+1}+ 
 2 (r + 2) (2 r^2 - 2 r - 9) a_{r+2} =0 \, ,
\fe
with $a_0=8$, $a_1=36$.   Given the values of $d_{r,r}$  all the other $d_{r, m}$ are determined by the Laplace difference equation \eqref{lapdiffSUN}.  
 
 In conclusion,  by summing both the perturbative \eqref{eq:pSU} and non-perturbative \eqref{eq:npSU} contributions in the $1/N$ expansion we find that the large-$N$ expansion of the $SU(N)$ integrated correlator \eqref{gsun} has the following structure, 
\ie
\cC_{SU(N)} (\tau, \bar \tau) = &\, {N^2 \over 4} +  \sum_{r=0}^{\infty} N^{{1\over 2}-r}  \sum^{\lfloor r/2 \rfloor}_{m=0} b_{r, m}E(\threeh {+} \delta_r {+} 2m; \tau, \bar \tau)    \\
&  \pm i    \sum_{r=0}^{\infty} N^{2-\frac{r}{2}} \sum_{m=0}^{r} d_{r,m} D_N\big(2m-\frac{3r}{2} ; \tau, \bar \tau\big) \, . 
\label{eq:genN}
\fe
This expression has to be understood as the formal yet complete large-$N$ transseries expansion of the integrated correlator $\cC_{SU(N)} (\tau, \bar \tau) $.
 The first line of \eqref{eq:genN} is an asymptotic series in the large-$N$ expansion, which was obtained in \cite{Chester:2019jas, Dorigoni:2021guq}. The second line gives the exponential corrections that are discussed in this paper and encoded in $B_{SU(N)}^{N\!P}(t)$. It should be stressed that the apparent $\pm i$ ambiguity in \eqref{eq:genN}, i.e. the jump in the Stokes constant, has to be understood from a resummation point of view. The first line of \eqref{eq:genN} is a formal asymptotic power series which can be resummed using \eqref{eq:I12adef}, while the seemingly ambiguous non-perturbative terms given by the second line of \eqref{eq:genN} can be resummed using \eqref{eq:I12def}. The sum of these two resummations produces our unambiguous starting equation \eqref{eq:ztomu}, i.e.
 \begin{align}
 \cC_{SU(N)} (\tau, \bar \tau)  &\notag =\cC^{P}_{SU(N)} (\tau, \bar \tau)  +\cC^{N\!P}_{SU(N)} (\tau, \bar \tau) \\
 &\notag =   2c_1(N)+2\!\!\!\sum_{(m,n)\neq(0,0)} \int_1^\infty e^{-t Y_{mn}(\tau,\bar\tau)} B_{SU(N)}^{P}(t) dt + 2\!\!\!\sum_{(m,n)\neq(0,0)} \int_1^\infty e^{-t Y_{mn}(\tau,\bar\tau)} B_{SU(N)}^{N\!P}(t) dt\\
 & = 2\!\!\! \sum_{(m,n)\neq(0,0)} \int_1^\infty e^{-t Y_{mn}(\tau,\bar\tau)} B_{SU(N)} (t) dt\,.
 \end{align}
 Note that although the transseries has been obtained at large-$N$, its Borel-Ecalle resummation does indeed provide a well-defined analytic continuation for all values of $N$ in the complex wedge $\mbox{Re}\,N>0$, and, in particular, it does coincide with equation \eqref{eq:geninv} for finite $N\in\mathbb{N}$.

Finally, it would be interesting to re-derive the non-perturbative corrections in \eqref{eq:genN} from the large-$N$ expansion of the spectral decomposition of \eqref{gsun} discussed in \cite{Collier:2022emf}. Presumably, the large-$N$ expansion of the spectral overlaps $\{\mathcal{C}_{SU(N)}, E_s\}$ of the integrated correlator \eqref{gsun} with the Eisenstein series $E_s = E(s;\tau,\bar\tau)$,  contains terms which are exponentially suppressed in $N$ and are responsible for the novel modular invariant functions $D_N(s;\tau,\bar\tau)$.

 \subsection{Holographic interpretation}
 \label{sec:largeNSUN}

  We will now  briefly discuss the holographic interpretation of the terms that are exponentially suppressed in the large-$N$ limit.  This is the large-$N$ limit in which contributions of Yang--Mills instantons, which are of order  $e^{-2\pi k \tau_2}$, are not suppressed, whereas they are exponentially suppressed  in $N$ in the 't Hooft limit.  Such  contributions arise in the non-zero Fourier modes of the Eisenstein series in the first line of  the expression for $\cC_{SU(N)}(\tau,\bar\tau)$  in \eqref{eq:genN}, and are dual to the contributions of  D-instantons to terms in the low energy expansion of the  holographically dual string theory.  
   
   Contributions to the integrated correlator in the large-$N$ limit with fixed $g_{_{Y\!M}}^2$ of the form \eqref{eq:modterms} are exponentially suppressed in the large-$N$ limit.  The holographic string theory interpretation  of  such contributions uses the identifications  $g_{_{Y\!M}}^2  = 4 \pi g_s= 4\pi/\tau_2$ and $\sqrt{ g_{_{Y\!M}}^2 N}= L^2/\alpha'$, where $g_s$ is the string coupling constant, $\alpha'$ is the square of the string length scale, and $L$ is the scale of the $AdS_5\times S^5$ space \cite{Maldacena:1997re,Gubser:1998bc,Witten:1998qj}.  Therefore, the large-$N$ expansion of the correlators with fixed $g_{_{Y\!M}}$ translates into the small-$\alpha'$ expansion of string amplitudes with fixed $g_s$. The existence of the terms that are exponentially suppressed in $N$  reflects the fact that the $\alpha'$ expansion of string amplitudes with fixed string coupling is an asymptotic series.  
   
     After these replacements the expression \eqref{eq:pqstrings}
 has the form of a sum over instanton contributions that correspond to  $\ell$ coincident euclidean world-sheets of $(p,q)$-strings (with ${\rm gcd}(p,q)=1$) wrapped on a two dimensional manifold of volume $L^2$.  Here we are identifying the tension of a $(p,q)$-string   \cite{Schwarz:1995dk,Aspinwall:1995fw}  with
\begin{equation}
T_{p,q} :=T_F {|p+q\tau|}  \,,
\label{eq:pqtense}
\end{equation}
where ${\rm gcd} (p,q)=1$ and $T_{1,0}=T_F:= 1/(2\pi \alpha')$ is the fundamental string tension.  Translating to string theory parameters, the exponential terms in   \eqref{eq:pqstrings}, or equivalently \eqref{eq:npSU}, become
\bea
\sum_{\ell=1}^\infty \sum_{{\rm gcd}(p,q)=1} \exp\Big( - 4\pi L^2 \ell \frac{|p+q\tau|}{2\pi \alpha'}  \Big) = \sum_{\ell=1}^\infty\sum_{{\rm gcd}(p,q)=1} \exp( - 4\pi L^2 \ell \,T_{p,q})  \,.
\label{eq:SUleadNP}
\eea 
 The sums in this expression  include a sum over multiple copies (indexed by $\ell$) of  euclidean $(p,q)$-string world-sheets. 
 When $g_s \ll1$  (i.e.~near the cusp $\tau_2\gg1$)  the fundamental string world-sheets dominate  while other $(p,q)$-string instantons dominate for other values of $g_s$ obtained by the appropriate action of $SL(2,\Z)$ on $\tau$.
 The complete exponentially suppressed contributions are given in \eqref{eq:genN} in a modular invariant form. 

We have not evaluated  the contribution of these instantons explicitly from string theory, but the factor of $4\pi L^2$ in the exponent in \eqref{eq:SUleadNP} suggests the contribution of $\ell$ coincident $(p,q)$-string euclidean world-sheets wrapping a great two-sphere, $S^2$, on the equator of the five-sphere, $S^5$.  
Although it is not obvious how such configurations would be stabilised, it is notable that their contribution to the integrated correlator \eqref{eq:genN} has an overall factor of $i$, which is characteristic of a negative fluctuation mode  (more generally, an odd number of negative modes).  Indeed, a two-sphere on the equator of the five-sphere would provide a saddle point that is  reminiscent, from a resurgence point of view, of uniton solutions in the principal chiral model \cite{Cherman:2013yfa,Cherman:2014ofa}. The semi-classical origin of such contributions certainly deserves further study.

Similarly, it would be interesting to develop a more detailed understanding of the holographic interpretation of  the saddle-point action $N A(\sqrt{Y_{mn}/4 N})$ that arose in \eqref{eq:modterms} with $A(x)$ defined in \eqref{eq:Action}.  As pointed out in \cite{Hatsuda:2022enx} the same function appeared in  the analysis   \cite{Drukker:2005kx} of  multiply-wrapped  Wilson loops in $\cN=4$ $SU(N)$ SYM in the 't Hooft limit, which is holographically described in  terms of a minimal surface  bordering the loop and embedded in $AdS_5$.  In that case  the argument of $A(x)$ was given by $x=\frac{k \sqrt{\lambda}}{4N}$ with $k$ being the winding number of a wound  Wilson loop.  According to  \cite{Drukker:2005kx}, such a multiply wound  Wilson loop can be effectively described by a four-dimensional embedded euclidean $D3$-brane carrying electric flux, with an action given by $N A( x)$. In the present context the holographic connection with the contribution of a $D3$-brane is hinted at by a naive application of the  AdS/CFT dictionary, which includes the identification $N= L^4/ (  4\pi  {\alpha'}^2 g_s)=2\pi^2 L^4 T_{D3}$, with $T_{D3}$ the $D3$-brane tension. This suggests that when $x$ is a fixed constant, the quantity $N A( x)$ should be identified with  the action of a euclidean $D3$ world-volume wrapped on a four-manifold.
 
We now turn to consider special large-$N$ limits in which the 't Hooft coupling $\lambda=g_{_{Y\!M}}^2 N$ is chosen as the independent coupling so $g_{_{Y\!M}}^2$ may depend on $N$.  In this way we will see how the non-perturbative results that were previously obtained by resurgence techniques in  \cite{Dorigoni:2021guq},  \cite{Collier:2022emf} and \cite{Hatsuda:2022enx} can be viewed as special limits of the $SL(2,\Z)$-invariant expression \eqref{eq:npSU}.

\section{Correspondence with resurgence results}
\label{sec:tHooft}

We will now consider the large-$N$ expansion in which  $\lambda = Ng^2_{_{Y\!M}} = 4\pi N/ \tau_2$ is an independent parameter in the range $1\ll  \lambda \ll   N$, which is the familiar strongly coupled 't Hooft limit.  In this case, the contributions of Yang--Mills instantons with instanton number $k$ are $O(e^{- 2\pi k N/\lambda})$.  In other words, in this regime Yang--Mills instantons, which  are present at every order in the $1/N$ expansion, are exponentially suppressed in $N$.  However, order by order in $1/N$, the large $\lambda$ perturbative expansion is not Borel summable, but can be completed via a resurgence argument. In  \cite{Dorigoni:2021guq} this argument was shown to give rise to further exponentially suppressed contributions of order $O(e^{-2\sqrt \lambda})$. 

Here we will see that this completion  follows very simply from the  $SL(2,\Z)$-invariant expression obtained in the previous section. 
%
For  $\lambda$ in the range $1\ll  \lambda \ll   N$ the  dominant contributions to the  exponentially suppressed terms  are those associated with the $m=\ell$, $n=0$  terms (the $
(\ell,0)$ terms) in \eqref{eq:npSU}.  In the holographic interpretation \eqref{eq:SUleadNP} these correspond to the contribution of $\ell$ coincident $(1,0)$-string (i.e.~fundamental string)  world-sheet instantons.  Since $q=0$ for these contributions they are independent of $\tau_1$ and only receive contributions from the zero mode of the integrated correlator, $\cC^{(0)}_{SU(N)}(\tau_2)$, which can be extracted from the $z^N$ term in \eqref{eq:0mode}.  As emphasised earlier, the $(\ell,0)$ terms contribute the first term in parentheses in  \eqref{eq:0mode}.   The non-perturbative contribution of this term is given by   \eqref{eq:npSU} upon substituting  $p=1$, $q=0$ into the definition of   $D_N(s;\tau,\bar\tau)$ in  \eqref{eq:Ddef}.  
The zero mode of this piece of $D_N(s;\tau,\bar\tau)$ is  $D_N^{(0),i}(s;\tau_2)$, which is defined in \eqref{eq:(i)}.

In this range of $\lambda$, we can rearrange these terms to take the form  
  \ie \label{eq:hordMT}
\cC^{N\!P,F}_{SU(N)}(\tau_2)= 4\sum_{\ell=1}^\infty \int_1^{\infty} e^{-\frac{\pi t \ell^2 }{\tau_2}}  B_{SU(N)}^{N\!P}(t)dt= \sum_{g=0}^\infty N^{2-2g}  \Delta\mathcal{C}^{(g)}(\lambda)\,,
\fe
where the superscript $F$ indicates that here we are considering only $F$-string (i.e.~$(1,0)$-string)  world-sheet  instantons. The functions $\Delta\mathcal{C}^{(g)}(\lambda)$ (denoted by $\pm i \Delta\mathcal{G}^{(g)}(\lambda)/2$ in \cite{Dorigoni:2021guq}) contain all the exponentially suppressed  large-$\lambda$ terms of the form
 \begin{equation}
 \Delta\mathcal{C}^{(g)}(\lambda) =\pm i \sum_{\ell=1}^\infty e^{-2\ell\sqrt{\lambda}} f_g(\ell\sqrt{\lambda})\,,\label{eq:CgSUMT}
 \end{equation}
 where $f_g(\ell\sqrt{\lambda})$ is a perturbative series in $1/\sqrt{\lambda}$.
 
The second term in parentheses in \eqref{eq:0mode} is the remaining contribution to the zero Fourier mode of the modular function  $\cC^{N\!P}_{SU(N)}(\tau,\bar\tau)$  appearing in \eqref{eq:npSU}.  This involves the zero-mode term $D_N^{(0),ii}(s;\tau_2)$ defined in \eqref{eq:(ii)}.  As emphasised before  \eqref{eq:0mode} this contribution is obtained by the zero mode of the sum over all values of $(m,n)$ with the exception of terms with $n=0$ or, in other words, by the zero mode of the infinite sum over all the multiple copies of $(p,q)$-strings with $q\neq0$.   We will now see that this contribution is proportional to $e^{-8\pi \ell N/\sqrt \lambda}$ with $\ell\in\mathbb{N}$ and $\ell\neq0$.

These remaining terms can be rewritten as  
\begin{equation} \label{eq:DDcontMT}
\cC^{N\!P,R}_{SU(N)}(\tau_2)  = 4 \sum_{\ell=1}^\infty  \int_1^\infty  e^{-\pi t \ell^2 \tau_2 }\frac{\sqrt{\tau_2}}{\sqrt{t}} \cB^{N\!P}_{SU}(z;t) dt = \sum_{g=0}^\infty N^{1-2g}  \Delta\tilde{\mathcal{C}}^{(g)}(\tilde{\lambda})\,,
\end{equation}
 where we have defined the ``dual'' 't Hooft coupling $\tilde{\lambda} =(4\pi N)^2 / \lambda$.
The superscript $R$ denotes the sum of the contribution of all terms that remain in the zero mode of the non-perturbative completion \eqref{eq:npSU} apart from the $(\ell,0)$ sector.

The functions $\Delta\tilde{\mathcal{C}}^{(g)}(\tilde{\lambda})$ (denoted $\pm i \Delta\tilde{G}^{(g)}(\tilde{\lambda})/2$ in \cite{Hatsuda:2022enx}) contain all the terms that are  exponentially suppressed in the ``dual'' 't Hooft coupling, which have the form $e^{-2\ell \sqrt{\tilde{\lambda}}}=e^{-8\pi \ell N/\sqrt \lambda}$ with $\ell \in \mathbb{N}$ and $\ell\neq0$.\footnote{The existence of such non-perturbative terms were first predicted in  \cite{Collier:2022emf}.}
By contrast with the $(1,0)$-string case, these remaining non-perturbative terms do not have a simple holographic interpretation since they arise from the  zero-mode contribution of the infinite sum over all the multiple copies (labelled by $\ell$) of $(p,q)$-strings with $q\neq0$. The contributions \eqref{eq:hordMT} to the zero mode were also found in \cite{Dorigoni:2021guq,  Hatsuda:2022enx} from  resurgence arguments at large-$\lambda$, while the terms \eqref{eq:DDcontMT} were found in \cite{Hatsuda:2022enx} from similar reasoning at large-$\tilde{\lambda}$, i.e.~$1\ll \tilde{\lambda}\ll N$ or equivalently $N\ll \lambda \ll N^2$.\footnote{The  exponential factor $e^{-2\sqrt{\tilde \lambda}}$ coincides with the leading factor contributing to a  $(0,1)$-string world-sheet instanton when $\tau_1=0$, This coincidence may be the reason that these contributions are referred to as ``$D$-string instantons'' in  \cite{Collier:2022emf, Hatsuda:2022enx}.}  

However, it should be stressed that the non-zero Fourier modes, which depend on $\tau_1$, cannot be determined easily by resurgence.  These contributions  are suppressed relative to  the $(1,0)$-string instanton contribution, $\cC^{N\!P,F}_{SU(N)}(\tau_2)$, but    they contribute with the same magnitude as $\cC^{N\!P,R}_{SU(N)}(\tau_2) \sim e^{-2 \sqrt{\tilde{\lambda}}}$ defined in \eqref{eq:DDcontMT}.   This is illustrated in \eqref{eq:(k)order}.

Up to now we have considered  approximations in which the saddle-point action $N A(x)$  has been expanded in small fluctuations of its argument \eqref{eq:Alargex}.   However,  in the  regimes $\lambda=O(N^2)$ (i.e.~$\tilde\lambda = O(1)$) and $\lambda=O(1)$ (i.e.~$\tilde\lambda = O(N^2)$)  the non-perturbative completion \eqref{eq:npSU} of the integrated correlator can be rearranged to produce different expansions.
We focus again on the zero-mode contributions discussed above and start by considering the non-perturbative $F$ terms in the regime $\lambda=O(N^2)$ (i.e.~$\tilde\lambda = O(1)$). This is the regime in which the on-shell saddle-point action $A(x)$ should not be expanded for $x$ small as in \eqref{eq:modeLarge}.
Schematically we obtain
\bea
\cC^{N\!P,F}_{SU(N)}(\tau_2) = \pm i \sum_{k=0}^\infty N^{2-k} \sum_{\ell=1}^\infty e^{-N A(\frac{\pi \ell}{\sqrt{\tilde{\lambda}}})} F_k(\frac{\pi \ell}{\sqrt{\tilde{\lambda}}}) \,, \qquad\qquad \lambda = O(N^2)\,,
\label{eq:electricMT}
\eea 
where the precise details of the expansion coefficients $F_k(x)$ are presented in \eqref{eq:electricD3b}.
 
A similar analysis applies to the remaining contribution to the zero Fourier mode when $\lambda=O(1)$ (i.e $\tilde \lambda = O(N^2)$).  In that case the non-perturbative $R$ contribution can be rearranged to give the expansion
\bea
\cC^{N\!P,R}_{SU(N)}(\tau_2) = \pm i\sum_{k=0}^\infty N^{\fiveh -k}\sum_{\ell=1}^\infty e^{-N A(\frac{\pi \ell}{\sqrt{{\lambda}}})} \tilde{F}_k(\frac{\pi \ell}{\sqrt{{\lambda}}}) \,,   \qquad\qquad  \lambda = O(1)\,,
\label{eq:magneticMT}
\eea
where details of the coefficients  $\tilde{F}_k(x)$ are presented in \eqref{eq:magneticD3b}.

In these regimes, the two zero-mode contributions to \eqref{eq:npSU} expanded as in \eqref{eq:electricMT} and \eqref{eq:magneticMT} become identical to the expressions in  \cite{Hatsuda:2022enx},  which were obtained from the asymptotic behaviour at large order of the large-$N$ genus expansion in the large-$\tilde{\lambda}$ or large-$\lambda$ regimes respectively, i.e.~from the large-$g$ behaviour of \eqref{eq:DDcontMT} and \eqref{eq:hordMT} respectively. 

In \cite{Hatsuda:2022enx}  the non-perturbative factors $e^{-N\!A(\ell \pi / \sqrt{\tilde{\lambda}})}$ and $e^{-N\!A(\ell \pi / \sqrt{{\lambda}})}$ in  \eqref{eq:electricMT} and \eqref{eq:magneticMT}  were called  ``electric'' and ``magnetic'' $D3$-brane instanton contributions in analogy with the results for multiply wound electric Wilson loop and magnetic 't Hooft loop derived in  \cite{Drukker:2005kx}. However, although the first zero-mode contribution  \eqref{eq:electricMT} does indeed come from the complete sum overl $\ell$ coincident $(1,0)$-string instantons, the second term arises as the zero-mode contribution from the infinite sum over $\ell$ coincident dyoonic $(p,q)$-string instantons with $q\neq0$.

Finally, more generally there are contributions from the non-zero modes that are also exponentially suppressed. In particular, the non-zero Fourier modes of $D_N(s; \tau, \bar \tau)$ are given in \eqref{eq:(ii)} and \eqref{eq:nzMode}.  These are the contributions from $(p, q)$-string instantons for which the exponential suppression is consistent with S-duality. We will not discuss these explicitly since their analysis involves a straightforward extension of the zero-mode analysis.

 \section{Integrated correlators with other classical gauge groups}
 \label{sec:gengroup}

The expression for  the $SU(N)$ integrated correlator, $\cC_{SU(N)}$, was extended to correlators for theories with general classical gauge groups $G_N=USp(2N)$ and $G_N=SO(n)$ (with $n=2N, 2N+1$) in \cite{Dorigoni:2022zcr}.    Recall that S-duality (Montinen--Olive duality) in the $SU(N)$ case   is generalised to Goddard--Nuyts--Olive  (GNO) duality  \cite{Goddard:1976qe} for general Lie groups.   
The correlators considered here are only sensitive to local  properties of S-duality and not to global features  that involve the centre of the gauge groups and their duals.   Such global properties are an essential feature of more general considerations, but here we only need to consider transformations associated with the Lie algebras.  These duality transformations correspond to the  following interchanges,
\bea
\mathfrak{su}(N) \leftrightarrow  \mathfrak{su}(N) \,, \qquad \mathfrak{so}(2N)  \leftrightarrow \mathfrak{so}(2N) \, , \qquad  \mathfrak{so}(2N+1)  \leftrightarrow \mathfrak{usp}(2N) \, , 
\label{eq:liegso}
\eea
which relate the  expressions for the integrated correlators in \cite{Dorigoni:2022zcr}   to each other, so we need only focus on $SO(n)$ gauge groups (in addition to the $SU(N)$ case described earlier).  Furthermore, recall that GNO duality implies invariance under the action of  $SL(2,\Z)$ on the complexified coupling $\tau$ in the correlators with $SU(N)$ and $SO(2N)$ gauge groups, while in the $SO(2N+1)$ and $USp(2N)$ cases the correlators are invariant under $\Gamma_0(2)$, which is a congruence subgroup of $SL(2,\Z)$. 

The corresponding generating functions are defined by
\ie
\label{eq:fsodef}
\cB^1_{SO} (z; t):= \sum_{n=1}^{\infty} B^1_{SO(n)}(t) z^n  \, , \qquad \cB^2_{SO} (z;t):= \sum_{n=1}^{\infty} B^2_{SO(n)}(t) z^n \, ,
\fe
where $B^1_{SO(n)}(t)$ and $B^2_{SO(n)}(t)$ are related to $\cC_{SO(n)}$ in \eqref{gsun}.
 In this section (and appendix \ref{app:clagroup}), we will derive the generating functions $\cB^1_{SO} (z; t)$ and $\cB^2_{SO} (z; t)$, and study the large-$N$ properties of the integrated correlators. 
 
  \subsection{Generating functions for other classical gauge groups}
   \label{sec:genSO}
  
 We begin by considering the generating function $\cB^2_{SO} (z; t)$. Recall, that  $B^2_{SO(2N)} (t)=0$, so we will focus on $SO(2N+1)$ gauge groups.  As shown in \cite{Dorigoni:2022zcr}, $B^2_{SO(2N+1)} (t)$ can be expressed as
\ie
\label{eq:B2int}
B^2_{SO(2N+1)} (t) = -{t \over 2} \int^{\infty}_0 e^{-xt} x^{3\over 2} \partial_x \left[ x^{3\over 2}  \partial_x  I^2_{SO(2N+1)}(x) \right] dx \, , 
\fe
where 
\ie
I^2_{SO(2N+1)}(x)  :=  e^{-x} \sum_{i=1}^N L_{2i-1}(2x) \, ,
\fe
and $L_{j}$ is the Laguerre polynomial. To proceed, we use the contour integral representation of the Laguerre polynomial, 
\ie
\label{eq:contL}
L_{j}(x) = \oint_{C} {e^{-x t_1/(1-t_1)} \over (1-t_1)  t_1^{j+1} } \frac{dt_1}{2\pi i}\, ,
\fe
where the contour $C$ circles the origin once in a counterclockwise direction. We can then perform the integration over $x$ in \eqref{eq:B2int}, which leads to
\ie
B^2_{SO(2N+1)} (t) =- \oint_{C}  \frac{3 t_1^{-2 N} \left( t_1^{2 N}-1\right) [t (t_1-1)+ t_1+1]}{2 [t_1
   (1-t)+t+1]^4} \frac{dt_1}{2\pi i}\, ,
\fe
and the generating function is given by  
 \ie
\cB^2_{SO} (z; t) := \sum_{N=1}^{\infty} B^2_{SO(2N+1)} (t) z^{2N+1} =\oint_{C}  \frac{3 \left(t_1^2-1\right) z^3 [t (t_1-1)+t_1+1]}{2 \left(z^2-1\right)  [t_1
   (1-t)+t+1]^4 \left(t_1^2-z^2\right)} \frac{dt_1}{2\pi i} \, . 
\fe
The relevant poles are located at $t_1=\pm z$, and the sum of their residues gives the final result
 \ie \label{eq:G2SO}
\cB^2_{SO} (z; t) = \frac{3 t (t+1) z^3 \left[ (t+1)^2 (3t^2-10t+3) -2 (t-1)^2 (t^2+10t+1)
   z^2 -(t-1)^4
   z^4\right]}{2
   \left[(t+1)^2-(t-1)^2
   z^2\right]^4} \, .
   \fe
As mentioned earlier, $B^2_{SO(n)} (t)$ vanishes for $n=2N$.  This is reflected in the above expression for  $\cB^2_{SO} (z; t)$, which is an odd function of $z$, and therefore  only receives contributions from   $SO(2N+1)$.
It will prove important that the singularity structure of $\cB^2_{SO} (z; t)$ is rather simple, with just two poles in $z$  located at
\ie
z = \pm \frac{(t+1)}{(t-1)}\,,
\label{eq:twopoles}
\fe
with equal and opposite residues since $\cB^2_{SO} (z; t) = -\cB^2_{SO} (-z; t)$.

We will now consider $\cB^1_{SO} (z; t) := \sum_{n=1}^{\infty} B^1_{SO(n)} (t) z^n $, where $B^1_{SO(n)} (t)$ is given by \cite{Dorigoni:2022zcr}
\ie \label{eq:B1SO}
B^1_{SO(n)} (t) = - {t\over 2} \int^{\infty}_0  e^{-xt} x^{3\over 2} \partial_x \left[ x^{3\over 2}  \partial_x  I^1_{SO(n)}(x) \right]dx \, , 
\fe
and
\ie \label{eq:ISO}
I^1_{SO(2N)}(x)&:=e^{-x}\,\sum_{i,j=1}^N \left( L_{2i-2} (x) L_{2j-2} (x) - L_{2i-2}^{2j-2i} (x) L_{2j-2}^{2i-2j} (x) \right) \,,\cr
I^1_{SO(2N+1)}(x)&:=e^{-x}\,\sum_{i,j=1}^N \left(  L_{2i-1} (x) L_{2j-1} (x) - L_{2i-1}^{2j-2i} (x) L_{2j-1}^{2i-2j} (x) \right)  \,.
\fe
Using combinations of the relations between sums of  bilinears of Laguerre polynomials given in  \eqref{eq:relations}  and making use of the contour integral representations of Laguerre polynomials leads to the generating function 
 \ie
\cB^1_{SO} (z; t) = \frac{1}{2} \cB_{SU} (z; t) + F_1(z;t) + t^{-1} F_1(z;t^{-1})+ F_2(z;t) + t^{-1} F_2(z;t^{-1})\, , \label{eq:soFull}
\fe
where $\cB_{SU} (z; t)$ is the generating function of the $SU(N)$ integrated correlator given in \eqref{eq:gresult}, and the functions $F_1(z;t)$ and  $F_2(z;t)$ are given by, 
\ie
F_1(z;t):= -\frac{3 t z^2 \left[ t
   (z-1)+2 (z+1) \right]}{2 x^4} \, ,
\fe
 and 
   \begin{align}
 &F_2(z;t) := {3z^2 \over 2 \left[(t+1)^2-(t-1)^2
   z\right]^{\frac{7}{2}} }\Big[ \frac{8   (z+1)^2 (1-z)^{5\over2}}{ x^4} 
   - 
   \frac{2
   \left(31 z^3+81 z^2+81
   z+31\right) (1-z)^{1\over2}}{ x^3}\\
   &+ \frac{  \left(211
   z^4+556 z^3+706 z^2+556
   z+211\right)}{(1-z)^{3\over2}  x^2}-\frac{7   \left(59
   z^5+123 z^4+138 z^3+138 z^2+123
   z+59\right)}{ (1-z)^{7\over2}   x}   \cr
 &- {1 \over 4z (1-z)^{7\over2}   }\Big( { x^5- 4 \left(7
   z+2 \right) x^4 + 2\left(106 z^2+125
   z+12\right) x^3 -2 \left(383
   z^3+797 z^2+451 z+17\right) x^2} 
  \cr  &\!\!\!\!\! + { \left(1583
   z^4+4286 z^3+4446 z^2+1758
   z+23\right) x- 2 
   \left(1013 z^5+2813 z^4+3752
   z^3+2888 z^2+1051 z+3\right)} \Big)\Big]\, ,\nonumber
   \end{align}
where  $x= t (1-z)+2 (z+1)$. 
Given the  above expressions, it is easy to see what the singularity structure of \eqref{eq:soFull} is.
  There are two poles located at
\ie
z= \frac{(t+2)}{(t-2)}\,,\qquad\qquad z=\frac{(1+2t)}{(1-2t)}\,,
\fe
as well as the same branch points as for the $SU(N)$ case, located at
\ie
z=1\,,\qquad\qquad z=z_1 = \Big(\frac{t+1}{t-1}\Big)^2\,.
\fe

It is straightforward to verify that $\cB^2_{SO} (z; t)$ satisfies the following homogenous differential equation,
\ie
t\, \partial_t^2 \left( t \cB^2_{SO} (z; t) \right) - {1\over 4}  \left[ z^2 \partial_z^2 \left( \cB^2_{SO} (z; t)  \left( z- z^{-1} \right)^2 \right)  \right]
 =0 \, ,
\fe 
while $\cB^1_{SO} (z; t)$ obeys an inhomogeneous differential equation  given by
\ie
t\, \partial_t^2 \left( t \cB^1_{SO} (z; t) \right) - {1\over 4}  \left[ z^2 \partial_z^2 \left( \cB^1_{SO} (z; t)  \left( z- z^{-1}\right)^2 \right)  \right]
- \left[ z^2 \partial_z  \left(z^{-1} \cB_{SU} (z; t) \right) -z \partial_z  \left( z \cB_{SU} (z; t) \right)  \right] =0 \, ,
\fe
where $\cB_{SU} (z; t)$ is the generating function of the $SU(N)$ correlator given in \eqref{eq:gresult}. These differential equations imply a Laplace difference equation for the integrated correlator\footnote{Note that for the special case $SO(3)$, the integrated correlator is given by $\C_{SO(3)}(\tau/2, \bar \tau/2)$ rather than $\C_{SO(3)}(\tau, \bar \tau)$ \cite{Dorigoni:2022zcr}.} $\C_{SO(n)}(\tau, \bar \tau)$ (with `source terms'  $\C_{SU(n)}(\tau, \bar \tau)$) as found in \cite{Dorigoni:2022zcr} 
\begin{align}
\Delta_\tau \C_{SO(n)}(\tau, \bar\tau)  -  2 c_{SO(n)} & \Big[ \C_{SO(n+2)}(\tau, \bar\tau) -2 \,\C_{SO(n)}(\tau, \bar\tau) +\C_{SO(n-2)} (\tau, \bar\tau)  \Big] \nn\\
& - n\, \C_{SU(n-1)} (\tau, \bar\tau) +(n-1)\, \C_{SU(n)} (\tau, \bar\tau) =0 \, ,\label{eq:LapSO}
\end{align} 
where the central charge is $c_{SO(n)} = n(n-1)/8$, and a similar equation for $USp(n)$ (with $n=2N$)
\begin{align}
\Delta_\tau \C_{USp(n)}(\tau, \bar\tau)  -  2 c_{USp(n)} & \Big[ \C_{USp(n+2)}(\tau, \bar\tau) -2 \,\C_{USp(n)}(\tau, \bar\tau) +\C_{USp(n-2)} (\tau, \bar\tau)  \Big] \nn\\
& + n\, \C_{SU(n+1)} (2\tau, 2\bar\tau) - (n+1)\, \C_{SU(n)} (2\tau, 2\bar\tau) =0 \, ,\label{eq:LapSp}
\end{align} 
with central charge $c_{USp(n)} = n(n+1)/8$.
The above Laplace difference equation uniquely determines $\C_{SO(n)}(\tau, \bar\tau)$ and $\C_{USp(n)}(\tau, \bar\tau) $ for any $n$ in terms of $\C_{SU(2)} (\tau, \bar\tau)$ \cite{Dorigoni:2022zcr}. Just as we argued previously in section \ref{sec:matrix0} for the $SU(N)$ case, this ultimately leads to the expression \eqref{gsun} for the integrated correlator with $SO(n)$ and $USp(2N)$ gauge groups.  
 
\subsection{Large-$N$ expansion for other classical gauge groups} 

It is well known that $\cN=4$ SYM with $SO(n)$ gauge group is dual to type IIB string theories in an orientifold with background $AdS_5 \times (S^5/\mathbb{Z}_2)$ \cite{Elitzur:1998ju, Witten:1998xy}. In considering the large-$n$ expansion of the integrated correlators  with $SO(n)$ gauge group, it was seen in \cite{Dorigoni:2022zcr} that  it is important to consider  the expansion in inverse powers of the combination
\ie
\tN := {1\over 4} (2n-1) \, ,
\fe
which is the total RR flux in the holographic dual theory, and we have $(L/\ell_s)^4 =2 g_{_{Y\!M}}^2 \tN$ (where $\ell_s$ is the string length scale). 
 It is easy to see that the contribution from $B^2_{SO(n)}(t)$ decays exponentially in the large-$\tN$ limit \cite{Dorigoni:2022zcr}.  Powers of $1/ \tilde N$ arise only from the large-$\tilde N$ expansion of  $B^1_{SO(n)}(t)$, and were obtained in \cite{Dorigoni:2022zcr} (see also  \cite{Alday:2021vfb}). They take exactly the same form as in the  large-$N$ expansion of  the $SU(N)$ correlator \eqref{eq:pSU} but with $N$ replaced by  $2\tN$, 
\ie \label{eq:genNso}
2\cC^{P}_{SO(n)} (\tau, \bar \tau) = \, {(2\tN)^2 \over 4} +  \sum_{r=0}^{\infty} (2\tN)^{{1\over 2}- r}  \sum^{\lfloor r/2 \rfloor}_{m=0} \tilde{b}_{r, m}E(\threeh {+} \delta_{r} {+} 2m; \tau, \bar \tau) \, ,
\fe
where, again, $\delta_r=0$ for even $r$ and $\delta_r=1$ for odd $r$ and the superscript $P$ indicates that these are the perturbative contributions in $1/\tN$.
 This expression can be derived from the generating function $\cB^1_{SO} (z; t)$ following the procedure given in section \ref{sec:largeN} for the $SU(N)$ case. The coefficients $\tilde{b}_{r, m}$ are also determined by the Laplace difference equation \eqref{eq:LapSO}, and some examples of them can be found in \cite{Dorigoni:2022zcr}; in fact $\tilde{b}_{r, \lfloor r/2 \rfloor}$ is identical to $b_{r, \lfloor r/2 \rfloor}$ of the $SU(N)$ correlator in \eqref{eq:genN}, as emphasised in \cite{Alday:2021vfb, Dorigoni:2022zcr}. 

We will now focus only on terms that decay exponentially and, since the analysis is very similar to that of the $SU(N)$ correlator, we will only list the results. Repeating the argument leading to \eqref{eq:GNint} from \eqref{eq:sunintcorr}, we proceed by closing the $z$ contour of integration, $C$, at infinity and collect the various contributions from the singular points of $\cB^i_{SO}(z;t)$ in the complex $z$-plane.
We begin with the contribution from $\cB^2_{SO} (z; t)$ given in \eqref{eq:G2SO}, which has two poles in $z$  
 located at $z = \pm \frac{(t+1)}{(t-1)}$ as shown in \eqref{eq:twopoles}.  In the large-$\tN$ expansion, this  leads to exponentially decaying terms of the form  
\ie \label{eq:soPeasy}
\cC^{2,N\!P}_{SO(n)} (\tau, \bar \tau) =     [1-(-1)^n]\frac{\sqrt{\pi}}{\sqrt{2}}   \sum_{r=0}^{\infty} \tN^{\frac{7}{4}-\frac{r}{2}} \sum_{m=0}^{r} {\tilde{d}}_{r, m} D_{\tN}\big(2m-\frac{3r}{2}-\frac{3}{4}; 2\tau, 2\bar \tau\big) \, ,
\fe
where $n=2N$ or $2N+1$ and the superscript $N\!P$ indicates that these are non-perturbative terms in $\tN$ at large $\tN$. We see that  $\cC^{2,N\!P}_{SO(2N)} (\tau, \bar \tau)$ vanishes as expected. 
Importantly, the non-holomorphic functions  that appear  in  $\cC^{2,N\!P}_{SO(2N+1)} (\tau, \bar \tau)$   have arguments $2\tau$ and $2\bar\tau$, which means that they are only invariant under the congruence subgroup $\Gamma_0(2)$ of $SL(2,\mathbb{Z})$, as expected from GNO duality.  Correspondingly, the exponential factor in $D_{\tN}(2m-\frac{3r}{2}-\frac{3}{4}; 2\tau, 2\bar \tau )$ has the form $\exp ( -2 \sqrt{2} \ell \sqrt{\frac{ \tilde N \pi }{\tau_2}} |p+2q\tau| )$ and so 
only the $(p,2q)$-string world-sheet instantons contribute.

 We have evaluated the   ${\tilde{d}}_{r,m}$ coefficients to very high orders.   The following are some low-order  examples, \ie\label{eq:easyd}
{\tilde{d}}_{0,0} &=2 \, , \cr
  {\tilde{d}}_{1,0} &=  -{9 \over 2^4}   \, , \quad \,\,\,\, {\tilde{d}}_{1,1} =   -{1\over 3} \, ,  \cr
{\tilde{d}}_{2,0} &= - {39\over 2^{10}}  \, ,\quad  \,\,  {\tilde{d}}_{2,1} = {15\over 2^{5}}  \, , \quad\quad \,\,\,{\tilde{d}}_{2,2} = {1 \over 36}\, ,   \cr
{\tilde{d}}_{3,0} &={45 \over 2^{15}} \, , \quad\quad\, {\tilde{d}}_{3,1} = {357\over 2^{11}} \, ,\quad \quad {\tilde{d}}_{3,2} = -{37 \over 640}  \, , \quad {\tilde{d}}_{3,3} =   - {1 \over 648} \, .
\fe
It is straightforward to show that $\cC^{2,N\!P}_{SO(n)} (\tau, \bar \tau)$ obeys the homogenous Laplace difference equation,  
\begin{align} \label{eq:lapsoHo}
\Delta_\tau \C^{2,N\!P}_{SO(n)}(\tau, \bar\tau)  -  2 c_{SO(n)} & \Big[ \C^{2,N\!P}_{SO(n+2)}(\tau, \bar\tau) -2 \,\C^{2,N\!P}_{SO(n)}(\tau, \bar\tau) +\C^{2,N\!P}_{SO(n-2)} (\tau, \bar\tau)  \Big]  =0 \, ,
\end{align} 
which can be used to determine the coefficients ${\tilde{d}}_{r, m}$ once the leading coefficients ${\tilde{d}}_{r, r}$ are given. 

The second contribution  $\cB^1_{SO} (z; t)$ \eqref{eq:soFull} contains both poles located at $z = \frac{(t+2)}{(t-2)}$ and at $z=\frac{(1+2t)}{(1-2t)}$, as well as two branch points, located at $z=1$ and $z=z_1$ (recall $z_1=\frac{(t+1)^2}{(t-1)^2}$). 
Just as in the $SU(N)$ case, the expansion near the branch point $z=1$ produces the perturbative asymptotic expansion in $1/\tN$ with coefficients given by finite rational linear combinations of Eisenstein series of half-integer index, i.e.~\eqref{eq:genNso} previously derived in \cite{Alday:2021vfb,Dorigoni:2022zcr}.

The non-perturbative terms are captured by the expansions near the other singular points.
Let us begin with the pole contributions. From the first pole at $z = \frac{(t+2)}{(t-2)}$ we find
\ie \label{eq:soP1}
\cC^{1,N\!P}_{SO(n)} (\tau, \bar \tau)\Big\vert_{z\sim \frac{(t+2)}{(t-2)}} =   \sqrt{\pi} \sum_{r=0}^{\infty} \tN^{\frac{7}{4}-\frac{r}{2}} \sum_{m=0}^{r} {2^{2m-\frac{r}{2}+\frac{5}{4}}} {\tilde{d}}_{r, m} D_{\tN}\big(2m-\frac{3r}{2} {-} \frac{3}{4}; \tau, \bar \tau\big) \, , 
\fe
where interestingly the coefficients ${\tilde{d}}_{r,m}$ are precisely the one we found from $\cC^{2,N\!P}_{SO(n)} (2\tau, 2\bar \tau)$ as given in \eqref{eq:easyd}. From the second pole at $z = \frac{(1+2t)}{(1-2t)}$ we obtain
\ie \label{eq:soP2}
\cC^{1,N\!P}_{SO(n)} (\tau, \bar \tau)\Big\vert_{z \sim \frac{(1+2t)}{(1-2t)}} = (-1)^n \sqrt{\pi} \sum_{r=0}^{\infty} \tN^{\frac{7}{4}-\frac{r}{2}} \sum_{m=0}^{r} {2^{\frac{r}{2}-2m-\frac{5}{4}}} {\tilde{d}}_{r, m} D_{\frac{\tN}{2}}\big(2m-\frac{3r}{2} {-} \frac{3}{4}; \tau, \bar \tau\big) \, , 
\fe
where again the coefficients ${\tilde{d}}_{r,m}$ are given by \eqref{eq:easyd}.
Furthermore, just as in the case of $\C^{2,N\!P}_{SO(n)}(\tau, \bar\tau)$, the pole contributions given in \eqref{eq:soP1} and \eqref{eq:soP2} obey the homogenous Laplace difference equation \eqref{eq:lapsoHo}. 

Finally, just as in   \eqref{eq:bn2int}, in   the case of  $SU(N)$ gauge groups,  the discontinuity of the Borel transform \eqref{eq:soFull} along $z\in[1,z_1]$ has a branch-cut of its own along $z\in (z_1,\infty)$ and this contribution takes exactly the same form as for the $SU(N)$ correlator \eqref{eq:npSU}
\ie \label{eq:socuts}
2\, \cC^{1,N\!P}_{SO(n)} (\tau, \bar \tau) \Big\vert_{z\sim  z_1}= \pm i   \sum_{r=0}^{\infty} (2\tN)^{2-\frac{r}{2}} \sum_{m=0}^{r} d'_{r, m} D_{2\tN}\big(2m-\frac{3r}{2}; \tau, \bar \tau\big) \, ,
\fe
which obeys the inhomogenous Laplace difference equation \eqref{eq:LapSO}. A few examples of the coefficients are given by
\ie
d'_{0,0} &= -2  \, , \cr
 d'_{1,0} &= {7 \over 3 }  \, , \quad\quad \,\, d'_{1,1} =  - {9  \over 4}   \, ,  \cr
d'_{2,0} &= -{37 \over 36}  \, ,    \quad \,  d'_{2,1} =-{15  \over 8}  \, ,  \quad  \,\, \,d'_{2,2} = - { 117 \over 64} \, ,  \cr
d'_{3,0} &={887  \over 3240} \, ,  \quad d'_{3,1} ={293 \over 160 }\, , \quad\,\,\, \, d'_{3,2} =  -{35  \over 128}   \, , \quad d'_{3,3} =   - {489 \over 512}   \, .
\fe
 Furthermore the particular coefficients $d'_{r, r}$ appearing in the above equation are identical to the coefficients $d_{r, r}$ of $\cC^{N\!P}_{SU(N)}(\tau, \bar \tau)$ as given in \eqref{eq:npSU}-\eqref{eq:coeffdrm}. This phenomenon should be compared with the perturbative terms given in the second line of \eqref{eq:pSU} for $SU(N)$ and in \eqref{eq:genNso} for $SO(n)$, for which we have an analogous relation between the coefficients $\tilde{b}_{r, \lfloor r/2 \rfloor}$ and $b_{r, \lfloor r/2 \rfloor}$, namely $\tilde{b}_{r, \lfloor r/2 \rfloor}=b_{r, \lfloor r/2 \rfloor}$. 

Summing \eqref{eq:soP1}, \eqref{eq:soP2} and \eqref{eq:socuts}, we obtain the complete non-perturbative contributions:
\begin{equation}
\cC^{1,N\!P}_{SO(n)} (\tau, \bar \tau)  = \cC^{1,N\!P}_{SO(n)} (\tau, \bar \tau) \vert_{z\sim \frac{(t+2)}{(t-2)}} +\cC^{1,N\!P}_{SO(n)}(\tau, \bar \tau)\Big\vert_{z \sim \frac{(1+2t)}{(1-2t)}}  +\cC^{1,N\!P}_{SO(n)} (\tau, \bar \tau) \Big\vert_{z\sim  z_1}\,.
\end{equation}
The leading large-$\tN$ non-perturbative contributions to $\cC^{1,N\!P}_{SO(n)}$ come from \eqref{eq:soP2}, and are schematically of the form
\ie
\exp\Big( -2\ell \sqrt{\frac{2 \tilde N \pi }{\tau_2}} |p+q\tau|\Big) = \exp\Big( - 2\pi L^2 \ell \frac{|p+q\tau|}{2\pi \alpha'}  \Big) = \exp( - 2\pi L^2 \ell T_{p,q})  \,,
\label{eq:SOleadNP}
\fe
with $T_{p,q}$ the $(p,q)$-string tension defined in \eqref{eq:pqtense}, and where we have generalised the holographic dictionary   to the case of $SO(n)$, so that  $\tau_2 = 4\pi/g_{Y\!M}^2$ and $\sqrt{ 2 g_{_{Y\!M}}^2 \tilde N}= L^2/\alpha'$. We see that the exponent in \eqref{eq:SOleadNP}  is half that of the  $SU(N)$ result \eqref{eq:SUleadNP}.  This can be understood by recalling that the $SO(2N)$ and $SO(2N+1)$ (and $USp(2N)$) theories are holographic duals of the type IIB string in an orientifold background ${AdS}_5\times ({S}^5 / \mathbb{Z}_2)$ that emerges from the near horizon geometry of $N$ coincident parallel $D3$-branes coincident with a parallel orientifold 3-plane ($O3$-plane).
Hence,  just as the $SU(N)$ result \eqref{eq:SUleadNP} can be understood in terms of $\ell$  $(p,q)$-string world-sheets wrapping an equatorial $S^2$ inside $S^5$, \eqref{eq:SOleadNP} should correspond to  $\ell$  $(p,q)$-string world-sheets  wrapping a maximal $S^2$ inside $\rm{S}^5 / \mathbb{Z}_2$.
Given the explicit expressions \eqref{eq:soPeasy}, \eqref{eq:soP1} and \eqref{eq:soP2}, the semi-classical configurations responsible for such non-perturbative corrections should have different semi-classical origins.  They may be local minima of the action or saddle points with an odd or even number of negative eigenvalues associated with the one-loop determinants.

Starting from the preceding  large-$\tN$,  fixed  $\tau$ results we can  extract the large-$\tN$ limit with fixed $\lambda_{SO(n)} = 2g_{Y\!M}^2 \tN = 8 \pi \tilde{N}/\tau_2$.   The argument is similar to one we provided for the $SU(N)$ case in the preceding section and appendix \ref{sec:Fstring}.  The result is that the  leading exponential terms contributing to the saddle-point approximation to $\cC^{N\!P}_{SO(n)}=\cC^{1,N\!P}_{SO(n)}+\cC^{2,N\!P}_{SO(n)}$ are given by the zero-mode contribution $D_N^{(0),i}$ of equations  \eqref{eq:soPeasy}, \eqref{eq:soP1}, \eqref{eq:soP2} and \eqref{eq:socuts} and their leading behaviour takes the form  $e^{-\sqrt{\lambda_{SO(n)}}}$,  $\,e^{-\sqrt{2\lambda_{SO(n)}}}$,   $ \,e^{-\sqrt{\lambda_{SO(n)}}}$ and $e^{-2\sqrt{\lambda_{SO(n)}}}$, respectively.
 We can also  define the ``dual'' 't Hooft coupling  $\tilde{\lambda}_{SO(n)} = 4(4\pi \tN)^2 / \lambda_{SO(n)} = 8 \tN \pi \tau_2$, and  consider the contributions from $D_N^{(0),ii}$ in the large-$\tN$ limit with fixed  $\tilde{\lambda}_{SO(n)}$ as discussed in appendix \ref{sec:remaining}.  There are again four contributions, which are given by\footnote{Note that the non-perturbative term  $\cC^{2,N\!P}_{SO(n)} (\tau, \bar \tau)$ in 
 \eqref{eq:soPeasy} is the only term that is  not invariant under $SL(2,\Z)$,  although it is invariant under  $\Gamma_0(2)$.  The fact that 
  $\tilde{\lambda}_{SO(n)}$ is related to ${\lambda}_{SO(n)}$ by $\tau_2 \to 1/\tau_2$, which is a transformation not contained in $\Gamma_0(2)$, accounts for the fact that the large-$\lambda_{SO(n)}$ behaviour and the large-$\tilde{\lambda}_{SO(n)}$ behaviour of $\cC^{2,N\!P}_{SO(n)} (\tau, \bar \tau)$ are different.} 
$e^{-2\sqrt{{\tilde{\lambda}}_{SO(n)}}}$,  $e^{-\sqrt{2{\tilde{\lambda}}_{SO(n)}}}$, $e^{-\sqrt{{\tilde{\lambda}}_{SO(n)}}}$ and $e^{-2\sqrt{{\tilde{\lambda}}_{SO(n)}}}$, respectively.

As stressed earlier,  since these results were obtained starting from the  manifestly duality-invariant large-$\tN$ limit with fixed $\tau$,  these different non-perturbative corrections are just facets of the sum over $(p,q)$-strings when expanded in different corners of the double-scaling limit of the parameter space $(\tN,\tau)$.

\section*{Acknowledgements}

The authors would like to thank the Isaac Newton Institute of Mathematical Sciences for support 
and hospitality during the programme ``New connections in number theory and physics'' when 
work on this paper was undertaken. This work was supported by EPSRC Grant Number EP/R014604/1. We are  particularly grateful to  Stefano Cremonesi, Nick Dorey, Iñaki García Etxebarria, Axel Kleinschmidt,  Arkady Tseytlin, Don Zagier and Shun-Qing Zhang for useful discussions.
CW is supported by a Royal Society University Research Fellowship No.~UF160350. No new data were generated or analysed during this study.

\appendix

 \section{Hermitian matrix model and the integrated correlator}
 \label{app:matrix1}

We will here review a few  basic properties of correlators of the  $N\!\times\! N$  hermitian matrix model and their connections with the integrated correlator \cite{Chester:2019pvm}. Following \cite{Morozov:2009uy},\footnote{We are grateful to Matteo Beccaria and Arkady Tseytlin for pointing out this reference.} the connected  $m$-point correlation functions of the matrix model are defined by, 
\ie
K_{i_1, \ldots, i_m} (N):= \langle {\rm tr} \phi^{i_1} \cdots  {\rm tr} \phi^{i_m}  \rangle_{\rm conn} = \int  {\rm exp}\left( - {1\over 2} {\rm tr} \phi^2 \right)  {\rm tr} (\phi^{i_1}) \cdots  {\rm tr} (\phi^{i_m} )d\phi  - {\rm disconnected \,\,\, parts} \, ,
\fe
where the integration is over the space of $N \!\times\! N$ hermitian matrices, and the measure is normalised such that $\langle 1 \rangle =1$. One may introduce a partition function, 
\ie
Z_N(\{t_k\}) := \int  {\rm exp}\left( - {1\over 2} {\rm tr} \phi^2 + \sum_k t_k \phi^k \right)d\phi   \, ,
\fe
in terms of which $K_{i_1, \ldots, i_m} (N)$ is given by
\ie
K_{i_1, \ldots, i_m} (N) = {\partial^m \over \partial_{t_{i_1}} \ldots \partial_{t_{i_m}}}  \log Z_N(\{t_k\}) \Big\vert_{\{t_k\}\to\{0\}}  \, . 
 \fe
Following \cite{Zagier1986}, it is convenient to introduce a generating functions for $K_{i_1, \ldots, i_m} (N)$ of the form 
 \ie
 e_N(x_1, \ldots, x_n) := \sum_{i_1, \ldots, i_m=0}^{\infty} K_{i_1, \ldots, i_m} (N) {x_1^{i_1} \ldots x_m^{i_m} \over i_1! \ldots i_m!} \, .
 \fe
 It is known that $e_N$ obeys Toda equations \cite{Morozov:2009uy}. For example,  $e_N(x_1)$ and $ e_N(x_1, x_2)$ satisfy
 \ie \label{eq:rec}
 e_{N+1}(x_1) -2e_N(x_1) +e_{N-1}(x_1) &= {x_1^2 \over N} e_N(x_1)\, , \cr 
  e_{N+1}(x_1, x_2) -2e_N(x_1, x_2) +e_{N-1}(x_1, x_2) &= {(x_1+x_2)^2 \over N} e_N(x_1, x_2) - {x_1 x_2 \over N^2} e_N(x_1) e_N(x_2) \, ,
 \fe
 where the initial ($N{=}1$)  values are 
 \ie \label{eq:ini}
 e_1(x_1) = \exp\left( {x_1^2 \over 2}\right) \, , \qquad  e_1(x_1, x_2) = \exp\left( {x_1^2 +x_2^2 \over 2}\right) \left( e^{x_1 x_2} -1 \right) \, . 
 \fe
 
It is useful to introduce a generating function for the $N$-dependence of $e_N(x_1, \ldots, x_n)$ \cite{Alexandrov:2003pj, Morozov:2009uy} that is given by 
  \ie
 e(x_1, \ldots, x_n; z) := \sum_{N=0}^{\infty} e_N(x_1, \ldots, x_n)  z^N \, . 
 \fe
The one-point function $e_N(x_1)$ was first obtained by  Zagier and Harer \cite{Zagier1986}, and the generating function $e(x_1; z)$ is also known explicitly, \cite{Alexandrov:2003pj, Morozov:2009uy} 
 \ie \label{eq:ge1}
 e(x_1; z) = {z\over (1-z)^2} \exp \left(\frac{x_1^2}{2}  \frac{1+z} {1-z} \right) \, .
 \fe
The generating function for the two-point function $e(x_1, x_2; z)$ is given by an integral representation,
  \ie  \label{eq:ge2}
 e(x_1, x_2; z) &= {z\over (1-z)^2} \Big\{ \exp \left({z (x_1+x_2)^2 \over 1-z} \right) e_1(x_1, x_2) \cr
& -  \int_0^z \Big[ \oint \!\! \oint   \! {1 \over (t-u_1 u_2)^2} \exp\! \left({ (x_1+x_2)^2 (z-t) \over (1-z)(1-t)}  \!+\! {x_1^2 \over 2} {1+u_1 \over 1-u_1}\!+\! {x_2^2 \over 2} {1+u_2 \over 1-u_2}\right){du_1\over 2\pi i} {du_2\over 2\pi i} \Big] dt\Big\} \, .
 \fe
 The contour is around the poles at $u_1=0$ and $u_2=0$ after expanding $ e(x_1, x_2; z)$ as a polynomial in $z$. 

We  will now discuss the connection between the perturbative part of the integrated correlator and the matrix model correlators.   The perturbative contribution to the integrated correlator \eqref{gsun} of the $SU(N)$ theory can be expressed in the form
 \ie
\mathcal{C}_{SU(N)}^{pert} (y) &=-\int_0^\infty   \frac{\omega } {2\sinh^2\omega} y^2  \partial_y^2 \, I_{SU(N)} \left(\frac{\omega^2}{y } \right) d  \omega\, ,    \label{pertSUn} 
\fe
 where $y=\pi \tau_2$. Importantly, it is known from  \cite{Chester:2019pvm} that   $I_{SU(N)} \left(\frac{\omega^2}{y } \right)$ is related to the matrix model two-point and one-point functions  introduced above by,\footnote{The matrix model one- and two-point functions $e_N(x_1), e_N(x_1, x_2)$ also have interesting applications to circular Wilson loops in $\mathcal{N}=4$ SYM \cite{Beccaria:2020ykg, Beccaria:2021alk}.} 
  \ie \label{eq:INeN}
I_{SU(N)} \left(\frac{\omega^2}{y } \right)= e_N\left(i {w\over \sqrt{y}},  - i {w\over \sqrt{y} }\right) + e_N\left(i {w\over \sqrt{y}} \right)^2 \, ,
 \fe
 a relation that enters in \eqref{eq:matrixrels} in the main text.
Using  the recursion relations \eqref{eq:rec} for $e_N\left(i {w\over y} \right)$ and $e_N\left(i {w\over y},  - i {w\over y}\right)$ and the initial $N{=}1$ conditions \eqref{eq:ini} results in the expression, 
   \ie
I_{SU(N)} \left(\frac{\omega^2}{y } \right) =2\, e^{-\frac{\omega^2}{y }} \sum_{i=1}^N (N-i)\Bigg[L_{i-1}\left({ \frac{\omega^2}{y}}\right)L_{i}\left({ \frac{\omega^2}{y}}\right)  + L_{i-1}^{1}\left({ \frac{\omega^2}{y}}\right)L_{i}^{-1}\left({ \frac{\omega^2}{y}}\right)\Bigg] \, ,   \label{pertSUn2} 
\fe
where $L_i^j(x)$ is the generalised Laguerre polynomial.  Interestingly, the above expression for  $I_{SU(N)}$ has a simpler form than the expression that was previously determined in \cite{Chester:2019pvm,Chester:2020dja} (see, for example, equation (A.39) in \cite{Chester:2020dja}).

\section{Some properties of $D_N(s;\tau,\bar\tau)$}
\label{app:newmod}

In this appendix we will study some basic properties of the non-holomorphic modular invariant functions $D_N(s; \tau,\bar\tau)$, defined in \eqref{eq:Ddef}, which enter into the exponentially suppressed terms that complete the large-$N$ expansion.
Recall $D_N(s; \tau,\bar\tau)$ is defined as\footnote{Non-holomorphic modular invariant functions analogous to $D_N(s; \tau,\bar\tau)$ have recently appeared in another context \cite{Luo:2022tqy}.}
\begin{align}
D_N(s; \tau,\bar\tau) &\label{eq:Ddef2}:= \sum_{(m,n)\neq(0,0)}  \exp\Big(-4\sqrt{N Y_{mn}(\tau, \bar \tau)} \Big) Y_{mn}(\tau, \bar \tau)^{-s} \\
&\label{eq:Ddef2Poincare}=  \sum_{\ell=1}^\infty \sum_{{\rm gcd} (p,q)=1 } \exp\Big(- 4 \sqrt{N \pi} \ell \frac{|p+q\tau|}{\sqrt{\tau_2}} \Big) \frac{1}{\pi^s} \frac{\tau_2^s }{\ell^{2s} |p+q\tau|^{2s}}\, .
\end{align}
It follows from the second line of this equation, \eqref{eq:Ddef2Poincare}, that the function $D_N(s; \tau,\bar\tau)$  can be expressed as the Poincar\'e sum 
\begin{align}
D_N(s; \tau,\bar\tau) & \,\,= \!\! \sum_{\gamma\in B(\mathbb{Z}) \backslash SL(2,\Z)} d_N(s; \gamma\cdot \tau ,\gamma\cdot\tau) \,,\label{eq:PoincareSum}
\end{align}
where the seed function is given by 
\begin{equation}
d_N(s; \tau ,\bar\tau) := \Big(\frac{\tau_2 }{\pi}\Big)^s\, \mbox{Li}_{2s}\Big(e^{- 4 \sqrt{\frac{  N\pi}{\tau_2}} }\Big)\,,\label{eq:PoincareSeed}
\end{equation}
with $\mbox{Li}_s(x) = \sum_{\ell=1}^\infty x^\ell/ \ell^s$ denoting the polylogarithm function and $\tau_2 =\mbox{Im}\,\tau$.  This seed function satisfies the  periodicity relation  $d_N(s; \tau+n,\bar\tau+n)  = d_N(s; \tau,\bar\tau ) $ for all $n\in Z$.
Consequently the sum in  \eqref{eq:PoincareSum} is over $SL(2,\Z)$ 
\begin{equation}
\gamma = \left(\begin{matrix} a & b \\ c & d \end{matrix}\right) \in SL(2,\Z)\,,\qquad \qquad \gamma \cdot \tau = \frac{a \tau+b}{c \tau+d}\,,
\end{equation}
modulo the Borel stabiliser
\begin{equation}
B(\Z) := \left\lbrace \left( \begin{matrix} \pm1 & n \\ 0 & \pm1 \end{matrix} \right)\Big\vert \,n\in \Z\right\rbrace\subset SL(2,\Z)\,.
\end{equation}
It is well known that the coset space $ B(\mathbb{Z}) \backslash SL(2,\Z)$ is isomorphic to $\{(p,q)\in \Z^2 \,\vert\, \rm{gcd}(p,q)=1\}$ so that \eqref{eq:Ddef2Poincare} equals the Poincar\'e sum \eqref{eq:PoincareSum}.

It is straightforward to show that $D_N(s; \tau,\bar\tau)$ obeys the Laplace equation, 
\ie \label{eq:LapD}
\Delta_{\tau} D_N(s; \tau,\bar\tau) - s(s-1) D_N(s; \tau,\bar\tau) = (4s-3) \sqrt{N} D_N(s-\textstyle{\frac{1}{2}}; \tau,\bar\tau)  +4 N  D_N(s-1; \tau,\bar\tau) \, .
\fe
When $N=0$, $D_N(s; \tau,\bar\tau)$ reduces to the non-holomorphic Eisenstein series $E(s; \tau, \bar \tau)$ and the above differential equation reduces to
\ie
\Delta_{\tau} E(s; \tau,\bar\tau) - s(s-1) E(s; \tau,\bar\tau) =0\, ,
\fe 
which is the well-known Laplace eigenvalue equations for the non-holomorphic Eisenstein series. 
For $N>0$, the exponential part plays the r\^ole of a regulator, which ensures that the lattice sum \eqref{eq:Ddef2} is convergent for all $s$. 

We will now consider the Fourier mode decomposition 
\begin{align}
D_N(s; \tau,\bar\tau)  = \sum_{k \in \mathbb{Z}} e^{2\pi i k \tau_1} D^{(k)}_N(s; \tau_2)  \,. 
\end{align}
and focus on the zero mode, $D^{(0)}_N(s; \tau_2)$. 
There are standard methods (see e.g. \cite{Ahlen:2018wng, Dorigoni:2019yoq}), that allow us to derive the Fourier modes of a Poincar\'e sum \eqref{eq:PoincareSum} in terms of an integral transform of its seed function \eqref{eq:PoincareSeed},  but we will  follow a different route here.

To obtain the Fourier decomposition of \eqref{eq:Ddef2} we first separate the sum over $(m, n) \neq (0, 0)$ into two terms: 
\begin{itemize}
\item[(i)] The sum over $(m, 0)$ with $m\neq 0$; 
\item[(ii)]The sum over $(m, n\neq 0)$ for $m \in \mathbb{Z}$. 
\end{itemize}

Case {(i)} is straightforward, giving
\begin{align} \label{eq:(i)}
D^{(0), i}_N(s; \tau_2) &=\sum_{m\neq0,n=0} \int_{-\frac{1}{2}}^{\frac{1}{2}}  \exp\Big(-4\sqrt{N Y_{mn}(\tau, \bar \tau)} \Big) Y_{mn}(\tau, \bar \tau)^{-s}  d\tau_1 = 2 \sum_{m =1 }^\infty e^{-4 m \sqrt{\frac{N \pi}{\tau_2}}} \left( {\tau_2 \over \pi m^2 }\right)^{s}\, ,
\end{align}
since when $n=0$ the variable $Y_{mn}(\tau, \bar \tau)$ reduces to $\pi m^2/\tau_2$ which is independent of $\tau_1$.

In order to consider  case  {(ii)}, it is useful to eliminate the square root in \eqref{eq:Ddef2} by introducing an integral representation for $D_N(s; \tau,\bar\tau)$, 
\begin{align}
D_N(s; \tau,\bar\tau) =\!\!\! \sum_{(m,n)\neq(0,0)}  \int_0^\infty \!\!e^{-t \,Y_{mn}(\tau, \bar \tau)} \Big[\frac{t^{s-1}{}_1F_1(1-s;\frac{1}{2}\vert {-}\frac{4N}{t})}{\Gamma(s)}-\frac{4\sqrt{N} t^{s-\frac{3}{2}}{}_1F_1(\frac{3}{2}-s;\frac{3}{2}\vert {-}\frac{4N}{t})}{\Gamma(s-\frac{1}{2})}\Big]dt\, .
\end{align}
We can now use standard Poisson resummation to obtain the Fourier series for case (ii), which takes the form
\begin{align}
&D^{ii}_N(s; \tau,\bar\tau) =    \nn\\
&\nn\sqrt{\tau_2} \sum_{\hat m \in \mathbb{Z}, n\neq 0} e^{2\pi i \hat {m} n \tau_1} \int_0^\infty  e^{- \pi \tau_2 \left( {\hat m^2 \over t} + n^2 t\right)} \Big[\frac{t^{s-{3\over 2}}{}_1F_1(1-s;\frac{1}{2}\vert  {-}\frac{4N}{t})}{\Gamma(s)}-\frac{4\sqrt{N} t^{s-2}{}_1F_1(\frac{3}{2}-s;\frac{3}{2}\vert {-} \frac{4N}{t})}{\Gamma(s-\frac{1}{2})}\Big]dt\, .\nn \\ 
 \label{eq:(ii)n}
\end{align}
The zero mode is given by  setting $\hat m =0$, giving 
\begin{align} \label{eq:(ii)}
D^{(0), ii}_N(s; \tau_2) =2\sqrt{\tau_2} \sum_{n=1}^\infty \int_0^\infty  e^{- \pi \tau_2n^2 t} \Big[\frac{t^{s-{3\over 2}}{}_1F_1(1-s;\frac{1}{2}\vert  {-}\frac{4N}{t})}{\Gamma(s)}-\frac{4\sqrt{N} t^{s-2}{}_1F_1(\frac{3}{2}-s;\frac{3}{2}\vert {-} \frac{4N}{t})}{\Gamma(s-\frac{1}{2})}\Big]dt\, .
\end{align}
Note that this second contribution can equally well be obtained from the zero-mode contribution of the sum over all the remaining terms $(m, n)$ with $m,n \in \mathbb{Z}$ and $n\neq0$, i.e.
\begin{align}
&D^{(0), ii}_N(s; \tau_2) = \sum_{n\neq 0}\sum_{m\in \Z} \int_{-\frac{1}{2}}^{\frac{1}{2}}  \exp\Big(-4\sqrt{N Y_{mn}(\tau, \bar \tau)} \Big) Y_{mn}(\tau, \bar \tau)^{-s}  d\tau_1\,.
\end{align} 

As an example we can consider $s=0$ and perform the $t$ integral to obtain, 
\begin{align}
D^{(0), ii}_N(0; \tau_2) = 2\sum_{n=1}^\infty 2 n \tau_2 K_1(4 n \sqrt{ N \pi \tau_2})\, ,
\end{align}
which, in the large-$N$ limit  this is  $ O(e^{-4  \sqrt{\pi N \tau_2}})$.  More generally, using the asymptotic properties of ${}_1F_1$, we see that  
\ie \label{eq:(ii)order}
D^{(0), ii}_N(s; \tau_2) =   O(e^{-4  \sqrt{\pi N \tau_2}})\, .
\fe
The complete Fourier zero mode is given by
\begin{equation}
D^{(0)}_N(s; \tau_2) = D^{(0), i}_N(s; \tau_2)+D^{(0), ii}_N(s; \tau_2)\,.
\end{equation}
When expressed in terms of the 't Hooft coupling $\lambda$ and the ``dual'' coupling $\tilde{\lambda} = (4\pi N)^2/\lambda$, the exponential behaviours \eqref{eq:(i)} and \eqref{eq:(ii)} become
\ie \label{eq:12cases}
e^{-2 \sqrt{\lambda}}\qquad\qquad \mbox{and}\qquad\qquad e^{-8 \pi N/ \sqrt{ \lambda}} =e^{-2\sqrt{\tilde{\lambda}}}\,.
\fe

Finally, the $k$-th non-zero Fourier mode,  $D_N^{(k)}(s;\tau_2)$ , is determined by considering $k = \hat{m}n$ (with $\hat m,n \ne0)$    in \eqref{eq:(ii)n}. 
For example,  when $s=0$ we find
\begin{equation}\label{eq:nzMode}
 D_N^{(k)}(0;\tau_2)  = 2\sum_{n \vert k} 4 n^2 \tau_2 \sqrt{ \frac{N}{ 4n^2 N + k^2 \pi \tau_2} } K_1\Big( 2\sqrt{\pi \tau_2 (4n^2 N + k^2 \pi \tau_2) } \Big)\,,
\end{equation}
where the sum runs over the positive divisors $n$ of $k$. Here we have assumed $k>0$ since $D_N^{(k)}(s;\tau_2)  = D_N^{({-}k)}(s;\tau_2)$  given that $D_N(s;\tau,\bar\tau)$, defined in \eqref{eq:Ddef2},   is real-analytic.
At large $N$ and fixed $\tau$ we have that  
\ie \label{eq:(k)order}
D^{(k)}_N(s; \tau_2) =   O(e^{-4  \sqrt{\pi N \tau_2}})\, ,
\fe
which for fixed $\lambda$ in the regime $1\ll \lambda \ll N$ becomes $O(e^{-8 \pi N/ \sqrt{ \lambda}})=O(e^{-2 \sqrt{\tilde{\lambda}}})$,  so it is of the same order as $D^{(0), ii}_N(s;\tau_2)$.  

Finally, it is straightforward to enlarge the space of non-holomorphic  modular functions $D_N(s; \tau,\bar\tau)$ to the space of  modular forms with holomorphic and anti-holomorphic weights $(w,w')$, by acting on $D_N(s; \tau,\bar\tau)$ with appropriate covariant derivatives.  
 This is analogous to the construction of non-holomorphic Eisenstein modular forms  that entered in the expressions for maximal $U(1)_Y$-violating correlators considered in  \cite{Dorigoni:2021rdo}. In that case the relevant forms had weights $(w,-w)$.  We would therefore expect that the large-$N$ expansions in that reference should require a non-perturbative completion by a series of  weight $(w,-w)$  modular forms $D^{(w)}_N(s;\tau,\bar\tau)$.      Following   \cite{Dorigoni:2021rdo}, a weight $(w,-w)$ modular form $D^{(w)}_N(s;\tau,\bar\tau)$  is obtained by applying a chain of $w$ covariant derivatives of the form $ D^{(w)}_N(s;\tau,\bar\tau) = \cD_{w-1} \dots \cD_0 \, D_N(s; \tau,\bar\tau)$, 
where the covariant derivative acting on a weight $(w,w')$ form is defined by 
\ie
\mathcal{D}_w = i \left(\tau_2 {\partial \over \partial{\tau}} - i{w\over 2} \right) \, ,
\label{eq:covderiv}
\fe
and  transforms it into a $(w+1,w'-1)$ form.

\section{Saddle-point analysis of contributions to the zero mode}
\label{app:zeromode}

This appendix presents details of the saddle-point analysis of the contributions to the zero mode, 
\begin{equation}
\cC^{(0)}_{SU(N)}(\tau_2) = \int_{-\frac{1}{2}}^{\frac{1}{2}} \cC_{SU(N)}(\tau,\bar\tau)  d\tau_1\,,
\end{equation}
in the 't Hooft limit in which $N\to \infty$ with $1\ll \lambda \ll N$.   As remarked in the main text the expression for the zero mode of the generating function, \eqref{eq:0mode}, consists of two types of terms:
\begin{itemize}
\item[(i)] the sum over $m=\ell\in \Z$, $n=0$, which is holographically dual to $\ell$ coincident $(1,0)$-string  world-sheet instantons; 
\item[(ii)] the zero mode of the sum over  $m \in \Z$ and $n=\ell  \ne 0 \in \Z$, i.e.~the zero mode of the infinite sum over all the multiple copies (labelled by $\ell$) of $(p,q)$-strings with $q\neq0$.  This is equivalent to setting $\hat m=0$ and summing over $n=\ell$ contributions, where $\hat m$ is the integer conjugate to $m$ in the Poisson summation.
\end{itemize}
 We stress that, unlike the $(1,0)$-string sector (i), the zero-mode contribution (ii) does not have a simple holographic interpretation although it was called the D-string instanton sector in  \cite{Collier:2022emf, Hatsuda:2022enx}.  Since the D-string is usually defined to be the $(0,1)$-string and depends on $\tau_1$ in an essential way \eqref{eq:SUleadNP}, this designation does not seem appropriate.

\subsection{The  $(1,0)$-string world-sheet instanton contribution}
\label{sec:Fstring}
 
 We now turn to details of the exponentially suppressed behaviour of the first non-constant term in the zero-mode integral \eqref{eq:0mode}, which corresponds to the contribution of the $(1,0)$-string (or $F$-string) world-sheet instanton.
In  order to consider the  large-$N$ contributions  we need to consider the saddle-point contribution to the contour integral
 \begin{align}
\cC^{N\!P,F}_{SU(N)}(\tau_2)& := 4\sum_{\ell=1}^\infty \int_1^{\infty} e^{-\frac{\pi t \ell^2 }{\tau_2}}  B_{SU(N)}^{N\!P}(t)dt\nn
\\
&\label{eq:FDcont}  = 4\sum_{\ell=1}^\infty \int_1^{\infty} e^{-\frac{\pi t \ell^2 }{\tau_2}}\Big[ \int_{z_1}^{\infty \pm i\epsilon} \frac{\mbox{Disc} \,\cB_{SU}(z;t)}{z^{N+1}}  \frac{dz}{2\pi i }\Big]dt\,,
\end{align}
where the superscript $F$ denotes the contribution of the $F$-string  instantons. We will see  that the large-$N$ expansion of this contribution at large 't Hooft coupling $\lambda$, produces a genus expansion $N^{2-2g}$, with $g\geq0$, of exponentially suppressed corrections at large $\lambda$.  These were identified in \cite{Dorigoni:2021guq} by applying resurgence analysis to the asymptotic large-$\lambda$ perturbation expansion of the genus expansion.

 The large-$N$, large-$\lambda$ expansion of $\cC^{N\!P,F}_{SU(N)}(\tau,\bar\tau)$  is controlled by a saddle point 
which, in the regime $1\ll \lambda \ll N$, is located at
 \begin{equation}
 t = t_1^\star = \frac{\sqrt{16 N^2 + \ell^2 \lambda}}{|\ell| \sqrt{\lambda}} \sim \frac{4 N}{|\ell| \sqrt{\lambda}} +... 
 \,,\label{eq:saddleLambda}
 \end{equation}
 and with an exponentiated  action given by  \ie
 e^{-S^{F}\!(t_1^*)}=\exp\Big[-\frac{\pi t^\star_1 \ell^2 }{\tau_2} - 2N \log\Big( \frac{t^\star_1+1}{t^\star_1-1} \Big) \Big]&= \exp\Big[-N A\Big(|\ell | \sqrt{\frac{\pi}{4 N\tau_2}} \Big)\Big] \\
 & = \exp\Big[-N A\Big(\frac{|\ell| \sqrt{\lambda}}{4N} \Big)\Big]  
\sim \exp\Big(  -2 |\ell| \sqrt{\lambda }\Big) \,,
\label{eq:electric}
 \fe  
where we have substituted the saddle-point action $NA(x)$ that was defined in \eqref{eq:Action}.
Note that this is precisely the exponential of the on-shell action \eqref{eq:Sos} evaluated for $(m,n)=(\ell,0)$.  In the second line we have further used the definition of the 't Hooft coupling and  kept the leading term of $A(x) = 8x +O(x^3)$ in the large-$N$ limit under consideration. 

 In order to consider the fluctuations around the saddle point we write  $ t = t_1^\star + \frac{N}{\lambda^{3/4}} \delta$ and expand the saddle-point action \eqref{eq:electric} in powers of $\delta$, and using the expression  \eqref{eq:gresult}  for $B_{SU}(z;t)$  into \eqref{eq:FDcont}  and the integration over $\delta$ leads to the final expression order by order in the $1/N$ expansion. 
At leading order, we obtain
 \begin{align}
N^2\Delta\mathcal{C}^{(0)}(\lambda) & \label{eq:C0}=  \pm\frac{ i }{2}   N^2 \sum_{\ell=1}^\infty e^{-2 \ell \sqrt{\lambda}} \Big[ 8 +\frac{18}{\ell\sqrt{\lambda}} +\frac{117}{4 \ell^2 \lambda}+\frac{489}{16 \ell^3 \lambda^{\frac{3}{2}}}+ \frac{3915 }{256 \ell^4 \lambda^2} +...\Big]\\
&\nn =  \pm\frac{ i }{2}   N^2 \sum_{\ell=1}^\infty  a_{\ell-1}(2 \sqrt{\lambda})^{1-\ell} \,\mbox{Li}_{\ell-1}(e^{-2\sqrt{\lambda}})\,.
 \end{align}
This is identical to the expression for $\pm i \Delta \mathcal{G}^{(0)}(\lambda)/2$ found in equation (5.39) in \cite{Dorigoni:2021guq} and the coefficients $a_{r}$, with $r\geq0$, are precisely related to the leading coefficients $d_{r,r} = -2^{-2(r+1)} a_r $ in \eqref{eq:npSU}.  According to the holographic correspondence such a term corresponds to a contribution to tree-level string theory associated with  $F$-string world-sheet instantons, and the higher order terms in $1/\sqrt{\lambda}$ correspond to the $\alpha'$-expansion in string theory.  It is straightforward to extend the above analysis and  determine the  next term in the $1/N$ expansion, which is a term of order  $N^0$. Following the same logic  as above we arrive at
 \begin{equation} \label{eq:C1}
N^0\Delta\mathcal{C}^{(1)}(\lambda) =  \pm \frac{i}{2} N^0 \sum_{\ell=1}^\infty e^{-2 \ell \sqrt{\lambda}} \Big[ -\frac{\ell^3 \lambda^{\frac{3}{2}}}{6}-\frac{3 \ell^2 \lambda}{8} + \frac{77 \ell \sqrt{\lambda}}{64}-\frac{127}{2^8}+\frac{927}{2^{12} \ell \sqrt{\lambda}} -\frac{3897}{2^{14} \ell^2\lambda}+...\Big]\,,
 \end{equation}
which  again agrees with the resurgence result $\pm i \Delta \mathcal{G}^{(1)}(\lambda)/2$ found in \cite{Dorigoni:2021guq}.\footnote{The first three terms in the parenthesis were missed in \cite{Dorigoni:2021guq}, as pointed out in  \cite{Hatsuda:2022enx}, see in particular equation (4.44) of  \cite{Hatsuda:2022enx}. }      
  
  We  therefore see that the $N^2$ and $N^0$ terms  in the large-$N$  expansion of the  zero mode of $\cC_{SU(N)}(\tau,\bar\tau)$  that are non-perturbative in $\lambda$  are consistent with a topological expansion of the form
  \ie \label{eq:hord}
 \cC^{N\!P,F}_{SU(N)}(\tau_2) = 4\sum_{\ell=1}^\infty \int_1^\infty e^{-\frac{ t \ell^2 \lambda}{4N}} \Big[ \int_{z_1}^{\infty\pm i \epsilon}  \frac{\mbox{Disc} \,\cB_{SU}(z;t)}{z^{N+1}}  \frac{dz}{2\pi i }  \Big]dt= \sum_{g=0}^\infty N^{2-2g}  \Delta\mathcal{C}^{(g)}(\lambda)\,,
\fe
 with $\Delta\mathcal{C}^{(g)}(\lambda)$ (denoted by $\pm i \Delta\mathcal{G}^{(g)}(\lambda)/2$ in \cite{Dorigoni:2021guq}) containing the exponentially suppressed  large-$\lambda$ terms of the form
 \begin{equation}
 \Delta\mathcal{C}^{(g)}(\lambda) =\pm i \sum_{\ell=1}^\infty e^{-2\ell\sqrt{\lambda}} f_g(\ell\sqrt{\lambda})\,,\label{eq:CgSU}
 \end{equation}
 where $f_g(\ell\sqrt{\lambda})$ is a perturbative series in $1/\sqrt{\lambda}$.   Combining these terms with the perturbative large-$\lambda$ expansion obtained from \eqref{eq:pSU} one obtains the transseries expansion 
 \begin{equation}  
  \cC_{SU(N)}(\tau,\bar\tau) \sim\, \mathcal{C}(\lambda) = \sum_{g=0}^\infty N^{2-2g}\big[ \mathcal{C}^{(g)}(\lambda) + \Delta\mathcal{C}^{(g)}(\lambda)\big]\,,\label{eq:genus}
 \end{equation}
 where all the non-perturbative contributions $\Delta\mathcal{C}^{(g)}(\lambda)$ can  be found from a resurgent analysis argument applied to the large-$\lambda$ expansion of $\mathcal{C}^{(g)}(\lambda)$ as discussed in \cite{Dorigoni:2021guq}, or equivalently using \eqref{eq:hord}. 
 Equation \eqref{eq:genus} ignores corrections that are exponentially suppressed in $N$ at large $N$, which will be discussed shortly.
 
 For fixed $\lambda$, the large-$N$ expansion of correlators corresponds to the genus expansion of string amplitudes.  Therefore the exact expression $A\Big(\frac{|\ell| \sqrt{\lambda}}{4N} \Big)$ for the on-shell action \eqref{eq:electric} can be interpreted as the result of resumming the genus expansion around the minimal surface. As discussed earlier, exactly the same function appears in the study of Wilson loops \cite{Drukker:2005kx}, and once again higher order terms in $A(x)$ can be thought as genus expansions around the minimal surface formed by the Wilson loop.  Furthermore, in the case of Wilson loops, the parameter $x$ is proportional to the electric charge $k$ of the Wilson loop, which may be tuned to scale with $N$. Therefore in the region where $x$ is not small, higher-order contributions to the expansion of $A(x)$ become important.  In this case  the string world-sheet  thickens  and an alternative description of $A(x)$ in terms of euclidean $D3$-brane instantons  is more appropriate  \cite{Drukker:2005kx}.\footnote{The present parameter $x$ is synonymous with the parameter $\kappa$ in the Wilson loop calculation in \cite{Drukker:2005kx}. } 
 
 This transition  between the small-$x$ and finite-$x$  descriptions is illuminated by expressing $x$ in terms of the  fundamental string (or $F$-string) and $D3$-brane tensions, 
\begin{equation}
x = \frac{|\ell| \sqrt{\lambda}}{4N}  = |\ell| \frac{T_F}{4 \pi L^2 T_{D3}}\,,\label{eq:xvar}
\end{equation}
where  $T_F = \sqrt{\lambda}/(2\pi L^2) \ll T_{D3}= N/(2\pi^2 L^4)$ is the fundamental string tension. 

We can perform a similar saddle-point analysis by considering the regime where the argument $x$ \eqref{eq:xvar} of the on-shell action \eqref{eq:electric} is kept constant in the large-$N$ limit.  This means considering the regime $\lambda = O(N^2)$ or equivalently the regime where the dual 't Hooft coupling $\tilde{\lambda}  = 4\pi N \tau_2= \frac{(4\pi N)^2 }{ \lambda}$ is kept fixed (i.e.~$\tilde \lambda = O(1)$) as $N$ becomes large.  In this case  $x = \frac{|\ell| \sqrt{\lambda}}{4N} = \pi |\ell| / \sqrt{\tilde{\lambda}}$ is also $O(1)$
  and the saddle point \eqref{eq:saddleLambda} is modified to 
\begin{equation}
t_1^\star = \frac{\sqrt{1+ x^2}}{ x}\,.\label{eq:elecSad}
\end{equation}
The fluctuations around the saddle point are obtained by writing $t=t_1^\star +N^{-\half} \delta $ and expanding the saddle-point action \eqref{eq:electric} in powers of $\delta$
\begin{equation}
S^F\!(t_1^\star+ N^{-\half} \delta) = NA(x)+ 4x^3\sqrt{1+x^2} \delta^2 +O( N^{-\half} \delta^3)\,.\label{eq:elecExp}
\end{equation} 
Upon expanding both the effective action and the integrand of \eqref{eq:electric} at large-$N$, or equivalently small $\delta$, and performing gaussian integrals over $\delta$, we arrive at what can be called the ``electric'' $D3$-brane expansion:
\begin{align}
 \cC^{N\!P,F}_{SU(N)}(\tau_2) &\label{eq:electricD3a} = \sum_{\ell=1}^\infty G^{\rm{(ele)}}\Big(N,\frac{\pi \ell}{\sqrt{\tilde{\lambda}}}\Big)\,,\\ 
 G^{\rm{(ele)}}(N, x) &\label{eq:electricD3b}  :=\pm  8 i\, e^{-N \!A(x)} \sum_{k=0}^\infty N^{2-k} \frac{h_k(x)}{[8 x(1+x^2)^\threeh]^k}\,.
\end{align}
 The expressions \eqref{eq:electricD3a}-\eqref{eq:electricD3b}  coincide with the results  of \cite{Hatsuda:2022enx} (where $x$ was called $y$).    In particular the coefficients $h_k({x})$, which are polynomials of order $4k$ in ${x}$, were presented in \cite{Hatsuda:2022enx} for $k\leq 3$. Higher-order polynomials can be determined straightforwardly from the saddle-point expansion.   For example, the $k=4$ term is given by
 \begin{align}
\!\!\!\!\!\!\!\!\!\! h_4(x) 
= -\frac{28256 x^{16}}{1215}-\frac{56512 x^{14}}{405}-\frac{10808 x^{12}}{45}-\frac{13664 x^{10}}{45}-\frac{549 x^8}{5}
+340 x^6+\frac{3185 x^4}{8}+\frac{1407 x^2}{8}+\frac{3915}{128}\,. 
 \label{eq:electricD3coeff} 
\end{align}
Higher order corrections, $h_{k\geq5}(x)$, can easily be computed from our saddle-point expansion.
 We stress that in the impressive analysis of \cite{Hatsuda:2022enx}, the equations \eqref{eq:electricD3a} and \eqref{eq:electricD3b}, together with the expressions for $h_{k\leq3}(x)$, were obtained from the asymptotic behaviour at large genus of the large-$N$ genus expansion in the large-$\tilde{\lambda}$ regime.
From our discussion, it is now manifest that the electric $D3$-instanton reduces to the world-sheet instanton in the 't Hooft limit.

 \subsection{The remaining zero-mode contributions}
\label{sec:remaining}
  
The second non-constant contribution to the zero mode  in \eqref{eq:0mode} is given by 
\begin{align}
\cC^{N\!P,R}_{SU(N)}(\tau_2) &\nn := 4 \sum_{\ell=1}^\infty  \int_1^\infty  e^{-\pi t \ell^2 \tau_2 }\frac{\sqrt{\tau_2}}{\sqrt{t}} \cB^{N\!P}_{SU}(z;t) dt\\
& \label{eq:DDcont} = 4\sum_{\ell=1}^\infty \int_1^{\infty} e^{-\pi t \ell^2 \tau_2 }\frac{\sqrt{\tau_2}}{\sqrt{t}} \Big[  \int_{z_1}^{\infty  \pm i\epsilon} \frac{\mbox{Disc} \,\cB_{SU}(z;t)}{z^{N+1}}  \frac{dz}{2\pi i } \Big]dt\,,
\end{align}
where the superscript $R$ denotes the remaining terms in the non-perturbative zero-mode contribution once the $(m,0)$  terms have been subtracted out.
As we have seen, this corresponds to the summand in \eqref{eq:Lattice1} after performing a Poisson summation that replaces  $m$ by $\hat{m}$ and then setting $(\hat{m},n)=(0,\ell)$.  
  
The leading factor in the  saddle-point analysis of this  contribution  is obtained by noting that the saddle-point  solution $t_1^*$ in \eqref{eq:saddles} now takes the form
 \begin{equation}
 t^\star_1 =  \frac{\sqrt{4 N+ \pi \ell^2 \tau_2 }}{\sqrt{\pi \ell^2 \tau_2 }} =  \frac{\sqrt{\lambda+\ell^2 \pi^2}}{|\ell|\pi}  \,.\label{eq:magsaddle}
 \end{equation}
 The exponentiated saddle-point action is given by
\ie
 e^{-S^{R}\!(t_1^*)}= \exp\Big[-\pi t^\star_1 \ell^2 \tau_2 - 2N& \log\Big( \frac{t^\star_1+1}{t^\star_1-1} \Big)\Big]   =  \exp\Big[-N A\Big(|\ell| \sqrt{\frac{\pi \tau_2}{4 N}} \Big)\Big]  \\
& = \exp\Big[-N A\Big(\frac{|\ell| \sqrt{\tilde{\lambda}}}{4N} \Big)\Big]  \sim   \exp\Big(- 2 |\ell| \sqrt{\tilde{\lambda}} \Big) \, ,
 \label{eq:magnetic}
\fe
where, following \cite{Collier:2022emf,Hatsuda:2022enx},  we have introduced the parameter $\tilde{\lambda} = (4\pi N)^2 / \lambda = 4\pi N \tau_2$ and consider the regime in which $1 \ll \tilde{\lambda}  \ll N$.

Higher order corrections can be  obtained straightforwardly resulting in the expansion
\begin{equation} \label{eq:DDcontApp}
\cC^{N\!P,R}_{SU(N)}(\tau_2)  = 4 \sum_{\ell=1}^\infty  \int_1^\infty  e^{-\pi t \ell^2 \tau_2 }\frac{\sqrt{\tau_2}}{\sqrt{t}} \cB^{N\!P}_{SU}(z;t) dt = \sum_{g=0}^\infty N^{1-2g}  \Delta\tilde{\mathcal{C}}^{(g)}(\tilde{\lambda})\,.
\end{equation}
The functions $\Delta\tilde{\mathcal{C}}^{(g)}(\lambda)$ are analogous to \eqref{eq:C0} and \eqref{eq:C1} and contain all the exponentially suppressed terms in the ``dual'' 't Hooft coupling of the form $e^{-2\ell \sqrt{\tilde{\lambda}}}=e^{-8\pi \ell N/\sqrt \lambda}$ with $\ell \in \mathbb{N}$ and $\ell\neq0$. In particular, these results precisely agree with the function $\Delta\tilde{\mathcal{C}}^{(g)}(\lambda)$ obtained in \cite{Hatsuda:2022enx}  (and denoted by $\pm i \Delta\tilde{G}^{(g)}(\tilde{\lambda})/2$ in this reference)  by resumming, order by order in $1/N$, the asymptotic large-$\tilde{\lambda}$ expansion using resurgent analysis.

 Just as in the discussion in section \ref{sec:Fstring}, we can consider  a saddle-point analysis in the regime where the argument  $\tilde{x} = \frac{|\ell| \sqrt{\tilde{\lambda}}}{4N} = \pi |\ell| / \sqrt{\lambda}$  of the saddle-point action \eqref{eq:magnetic} is kept $O(1)$ in the large-$N$ limit. This means that we are here considering the regime in which $\tilde{\lambda} = O(N^2)$, or equivalently, $\lambda  = O(1)$.  The saddle point \eqref{eq:magsaddle} can then be rewritten as
    \begin{equation}
    t_1^\star = \frac{\sqrt{1+\tilde{x}^2}}{\tilde{x}}\,,
    \end{equation} 
 and the quadratic fluctuations are obtained from the large-$N$ expansion, or equivalently small-$\delta$ expansion of the effective action
\begin{equation}
S^R\!(t_1^\star+ N^{-\half} \delta) = NA(\tilde{x})+ 4\tilde{x}^3\sqrt{1+\tilde{x}^2} \delta^2 +O( N^{-\half} \delta^3)\,.
\end{equation}
These expressions are similar to the ``electric'' results \eqref{eq:elecSad}-\eqref{eq:elecExp} upon exchanging $\tilde{x}\to x$. However, we see that  \eqref{eq:DDcont}   contains an important additional factor of  $\sqrt{\tau_2/t}$  resulting from the Poisson summation over $m$.
Expanding the saddle-point action  at large-$N$ and performing gaussian integrals over $\delta$, produces the expression 
\begin{align}
 \cC^{N\!P,R}_{SU(N)}(\tau_2) &\label{eq:magneticD3a} =  \sum_{\ell=1}^\infty \tilde{G}\Big( N,\frac{\pi \ell}{\sqrt{\lambda} }\Big)\,,\\ 
 \tilde{G}(N, \tilde{x}) &\label{eq:magneticD3b}  :=\pm16 i  \frac{ {\tilde{x} }^{\threeh}} {\sqrt{\pi } (1+\tilde{x}^2)^{\scriptsize{\frac{1}{4}}}} e^{-N A(\tilde{x})} \sum_{k=0}^\infty N^{\fiveh-k} \frac{g_k(\tilde{x})}{[8 \tilde{x}(1+\tilde{x}^2)^\threeh]^k}\, ,
\end{align}
which is a different expansion from the ``electric'' case  \eqref{eq:electricD3a}-\eqref{eq:electricD3b}.
 The expressions \eqref{eq:magneticD3a}-\eqref{eq:magneticD3b} again  coincide with the results of the \cite{Hatsuda:2022enx}  (modulo renaming $\tilde{x}$ by $x$ and $\tilde{G}$ by $G^{\rm{(mag)}}$).\footnote{
 In \cite{Hatsuda:2022enx} these non-perturbative terms $e^{-N\!A(\tilde{x})}$ were called ``magnetic $D3$-brane instantons''. However, since these terms arise as the zero mode of the infinite sum over all multiple copies (labelled by $\ell$) of $(p,q)$-string instantons with $q\neq0$, the nomenclature ``magnetic $D3$-brane'' does not seem appropriate.}
 The coefficients $g_k(\tilde{x})$ are polynomials of order $4k$ in $\tilde{x}$ that  were determined in \cite{Hatsuda:2022enx} for $k\leq 3$.  Higher-order terms can again be determined straightforwardly.  For example, the $k=4$ term is given by 
\begin{align} \label{eq:magneticD3coeff} 
&g_4(\tilde{x})   \\
&=-\frac{28256 \tilde{x}^{16}}{1215}-\frac{56512 \tilde{x}^{14}}{405}-\frac{32188 \tilde{x}^{12}}{135}-\frac{1504 \tilde{x}^{10}}{3}-\frac{102233 \tilde{x}^8}{240}+\frac{181 \tilde{x}^6}{16}+\frac{27909 \tilde{x}^4}{256}+\frac{1815 \tilde{x}^2}{32}+\frac{343035}{32768}\,.  \nn 
\end{align}

As in the ``electric'' case, in \cite{Hatsuda:2022enx}  these expressions were determined by analysis of the asymptotic behaviour of the large-$N$ genus expansion  at high genus in the large-$\lambda$ regime and they reduce to \eqref{eq:DDcontApp} in the regime $N\ll \lambda \ll N^2$, i.e.~the ``dual'' 't Hooft regime $1\ll \tilde{\lambda} \ll N$.

To conclude this section, we emphasise that the two distinct non-perturbative terms, \eqref{eq:electric} and \eqref{eq:magnetic},  are the two parts \eqref{eq:(i)} and \eqref{eq:(ii)} of the zero Fourier mode of the $SL(2, \mathbb{Z})$ invariant function $D_N(s; \tau, \bar \tau)$. Indeed, \eqref{eq:C0} and \eqref{eq:C1}  can be obtained directly from \eqref{eq:npSU} by replacing $D_N(s; \tau, \bar \tau)$ with its zero mode \eqref{eq:(i)}. 
Similarly, the expansion for the non-perturbative terms at large-$\tilde{\lambda}$ \eqref{eq:magnetic}, derived in \cite{Hatsuda:2022enx}, is recovered from \eqref{eq:npSU} by replacing $D_N(s; \tau, \bar \tau)$ with the remaining zero-mode contribution \eqref{eq:(ii)}.

Therefore, the sum  \eqref{eq:npSU}  contains all the non-perturbative terms obtained by resurgence at large-$\lambda$ and large-$\tilde \lambda$  in \cite{Dorigoni:2021bvj,Dorigoni:2021guq,Collier:2022emf,Hatsuda:2022enx}, and from resurgence at large $N$ in \cite{Hatsuda:2022enx}.  
We see that despite the fact that the manifest S-duality of \eqref{eq:npSU} is obscured in considering the different  large-$N$ 't Hooft limits of the $F$-string \eqref{eq:electric}-\eqref{eq:electricD3a} and of the zero mode of the sum over all remaining $(p,q)$-strings with $q\neq0$ \eqref{eq:magnetic}-\eqref{eq:magneticD3a}, these results contain  remnants of the relations implied by $SL(2,\Z)$.

\section{Generating Functions for general classical groups}
\label{app:clagroup}

This appendix  presents some of the details used in deriving the generating functions for the integrated correlators for theories with  general classical groups given in section \ref{sec:gengroup}.  The methods used in the $SU(N)$ case in appendix \ref{app:matrix1} and section \ref{sec:matrix0} do not generalise to general classical groups in an obvious manner so we will use a method that applies to all cases.

In order to evaluate the generating functions it is necessary to reduce the double sums in \eqref{eq:ISO} to single sums. This can be achieved by using a representation of Laguerre polynomials in terms of   creation and annihilation operators (as in  \cite{Okuyama:2018yep}), 
\ie \label{eq:caop}
\left\langle i\left|e^{\sqrt{x}\left(a+a^{\dagger}\right)}\right| j\right\rangle=\left\langle j\left|e^{\sqrt{x}\left(a+a^{\dagger}\right)}\right| i\right\rangle=\sqrt{\frac{i !}{j !}} e^{\frac{x}{2}}x^{\frac{j-i}{2}} L_{i}^{(j-i)}\left(-x\right)\, , 
\fe
where
\ie
\left[a, a^{\dagger}\right]&=1, \quad\qquad a|0\rangle=0, \quad \qquad|k\rangle=\frac{\left(a^{\dagger}\right)^{k}}{\sqrt{k !}}|0\rangle\, .
\fe
We begin by inserting the projector $(1 \pm (-1)^i)/2$ in \eqref{eq:ISO}, which leads to
\ie \label{eq;so}
I^1_{SO(2N)}(x)=\frac{1}{4}e^{-x}\sum_{i,j=0}^{2N-1} \left[1+2(-1)^i+(-1)^{i-j}\right]\left(L_i(x) L_j (x) - L_{i}^{j-i} (x) L_j^{i-j}(x)  \right) \, , \cr
I^1_{SO(2N+1)}(x)=\frac{1}{4}e^{-x}\sum_{i,j=0}^{2N-1} \left[1-2(-1)^i+(-1)^{i-j}\right]\left(L_i(x)  L_j (x) - L_{i}^{j-i} (x) L_j^{i-j}(x)  \right)  \, .
\fe
Each term in the above equations can be simplified by using  \eqref{eq:caop}. For example, 
\begin{align}
\sqrt{-x} e^{-x}\sum_{i,j=0}^{n-1}L_i^{j-i}(x) L_j^{i-j}(x)&=\sum_{i,j=0}^{N-1} \left\langle i\left|\sqrt{-x}\,e^{\sqrt{-x}\left(a+a^{\dagger}\right)}\right| j\right\rangle \left\langle j\left|e^{\sqrt{-x}\left(a+a^{\dagger}\right)}\right| i\right\rangle\\
&=\sum_{i,j=0}^{N-1} \left\langle i\left|\left[a,e^{\sqrt{-x}\left(a+a^{\dagger}\right)}\right]\right| j\right\rangle \left\langle j\left|e^{\sqrt{-x}\left(a+a^{\dagger}\right)}\right| i\right\rangle\cr 
&=\sum_{j=0}^{N-1} \left(\sum_{i=0}^{N-1}  \left\langle i+1\left|\,e^{\sqrt{-x}\left(a+a^{\dagger}\right)}\right| j\right\rangle \left\langle j\left|e^{\sqrt{-x}\left(a+a^{\dagger}\right)}\right| i\right\rangle-(i\rightarrow i-1)\right)\, , \nn
\end{align}
where we have utilised the symmetric property of the inner products and the following relation, 
\ie
\sqrt{x} e^{\sqrt{x}\left(a+a^{\dagger}\right)}&=\left[a, e^{\sqrt{x}\left(a+a^{\dagger}\right)}\right]=\left[e^{\sqrt{x}\left(a+a^{\dagger}\right)}, a^{\dagger}\right]\,.
\fe
This leads to an expression with one less summation index to be summed 
\ie
\sqrt{-x} e^{-x}\sum_{i,j=0}^{N-1}L_i^{j-i}(x) L_j^{i-j}(x)=e^{-x}\sum_{i=0}^{N-1}\frac{N}{\sqrt{-x}}L_{N}^{i-N}(x)L_{i}^{N-1-i}(x)\,.
\fe
Focussing now on the term with the alternating sign $(-1)^{i}$, 
\ie
e^{-x}\sum_{i,j=0}^{N-1} (-1)^{i} L_i^{j-i}(x) L_j^{i-j}(x)=\sum_{i,j=0}^{N-1} (-1)^{i} \left\langle i\left|e^{\sqrt{-x}\left(a+a^{\dagger}\right)}\right| j\right\rangle \left\langle j\left|e^{\sqrt{-x}\left(a+a^{\dagger}\right)}\right| i\right\rangle\, ,
\fe
we can consider the $x$ derivative of this double sum, 
\ie
-\partial_{x} \left(e^{-x}\sum_{i,j=0}^{N-1} (-1)^{i} L_i^{j-i}(x) L_j^{i-j}(x)\right)=& \sum_{i,j=0}^{N-1} \frac{(-1)^{i}}{\sqrt{-x}} \left\langle i\left|e^{\sqrt{-x}\left(a+a^{\dagger}\right)}\left(a+a^{\dagger}\right)\right| j\right\rangle \left\langle j\left|e^{\sqrt{-x}\left(a+a^{\dagger}\right)}\right| i\right\rangle\cr
=& \sum_{i,j=0}^{N-1}\frac{(-1)^{i}}{\sqrt{-x}} \Big[\sqrt{i}\left\langle i\left|e^{\sqrt{-x}\left(a+a^{\dagger}\right)}\right| j\right\rangle\left\langle j\left|e^{\sqrt{-x}\left(a+a^{\dagger}\right)}\right| i-1\right\rangle \cr
&+ \sqrt{i+1}\left\langle i\left|e^{\sqrt{-x}\left(a+a^{\dagger}\right)}\right| j\right\rangle\left\langle j\left|e^{\sqrt{-x}\left(a+a^{\dagger}\right)}\right| i+1\right\rangle \Big] \, .
\fe
The sum over $i$ is now straightforward and  leads to the expression,  
\ie
\partial_{x} \left(e^{-x}\sum_{i,j=0}^{N-1} (-1)^{i} L_i^{j-i}(x) L_j^{i-j}(x)\right)=
e^{-x}\sum_{i=0}^{N-1} (-1)^{N-1} \frac{N}{x} L_{N}^{i-N}(x)L_{i}^{N-1-i}(x)\,.
\fe
Similarly, we have
\begin{align}
& e^{-x}  \sum_{i,j=0}^{N-1} L_{2i}(x) L_{2j}(x) = \frac{1}{4}\,e^{-x}\sum_{i,j=0}^{2N-1}\left(L_i(x)  L_j(x)  +2\,(-1)^iL_i(x)  L_j(x)  +(-1)^{i-j}L_i(x)  L_j(x) \right)\cr
&=\frac{1}{4}\left[ e^{-x}\left(L_{2N-1}^1(x) \right)^2-2e^{-x/2}\,L^1_{2N-1}(x) \int^x \frac{e^{-x'/2}}{2}L_{2N-1}^1(x') dx'+\Big(\int^x\frac{e^{x'/2}}{2}L^1_{2N-1}(x') dx'\Big)^2\right] \cr
&=\frac{1}{4}\left[\int^x e^{-x'/2} L_{2n-2}^2 (x')\,dx' \right]^2\,,
\end{align}
where we  have used the recurrence relation for Laguerre polynomials $L^\alpha_n=L^{\alpha+1}_n-L^{\alpha+1}_{n-1}$, and completed the square.

The above considerations lead to the following  simplified relations that are useful for evaluating generating functions in the main text,
\ie \label{eq:relations}
e^{-x}\sum_{i,j=0}^{N-1}  L_{2i +\delta} (x) L_{2j +\delta}(x) &={1\over 4} \left[\int^x e^{-x'/2} L_{2N-2 +\delta}^2 (x')\,dx' \right]^2\, , \\
 e^{-x}\sum_{i,j=0}^{N-1}  L_{i}^{j-i} (x) L_{j}^{i-j} (x) &= e^{-x}\sum_{i=0}^{N-1} \frac{N}{x} L_{i}^{N-1-i} (x) L_{N}^{i-N}(x) \, , \\
e^{-x}\sum_{i,j=0}^{N-1} (-1)^i L_{i}^{j-i}(x)  L_{j}^{i-j} (x) &= \int^x e^{-x'} \sum_{i=0}^{N-1} (-1)^{N-1} \frac{N}{x'} L_{i}^{N-1-i} (x') L_{N}^{i-N}(x') \, dx'\, , \\
e^{-x}\sum_{i,j=0}^{N-1} (-1)^{i-j} L_{i}^{j-i}(x)  L_{j}^{i-j} (x) &= \int^x e^{-x'} \sum_{i=0}^{N-1} (-1)^{N-1+i} \frac{N}{x'} L_{i}^{N-1-i} (x') L_{N}^{i-N} (x') \,dx' \, ,
\fe
where $\delta$ is either $0$ or $1$.   In order to determine the generating function $\cB^1_{SO}(z;t)$, we need to evaluate the integral \eqref{eq:B1SO} for each term given in \eqref{eq;so}.  For example, for the last term in \eqref{eq:relations}, we have to compute
\ie \label{eq:examp1}
\sum_{N=1}^{\infty} -t\int_0^\infty  e^{-tx}\,\frac{x^{\frac{3}{2}}}{2} \partial_{x}\left\{x^{\frac{3}{2}} \partial_{x}\int^x e^{-x'} \sum_{i=0}^{N-1} (-1)^{N-1+i} \frac{N}{x'} L_{i}^{N-1-i} (x') L_{N}^{i-N} (x') \,dx' \right\} z^N  dx\, .
\fe
We start by making use of the last expression in  \eqref{eq:relations} and the relation
\ie
L_i^{N-1-i}(x) L_N^{i-N}(x)=\oint\!\!  \oint  \frac{\exp \left(-x\left(\frac{t_1}{1-t_1}+\frac{t_2}{1-t_2}\right)\right)}{(1-t_1)^{N-i} t_1^{i+1}(1-t_2)^{i-N+1}t_2^{N+1}} \frac{dt_1}{2\pi i}\frac{dt_2}{2\pi i} \, ,
\fe
 which follows from the contour integral representation of Laguerre polynomials \eqref{eq:contL}. 
The summations over $i$ and $N$ and the integration over $x$ in \eqref{eq:examp1} are all elementary, leading to
\ie
 \oint\!\!  \oint  dt_1 dt_2\frac{t (t_1-1)^2 (t_2-1)^2 z (t (t_1-1) (t_2-1)+3 t_1 t_2-3) \left(t_1^2
   (t_2-1) t_2+t_1 z^2-z^2\right)}{4 (t (t_1-1) (t_2-1)-t_1 t_2+1)^3
   (z-t_1 t_2)^2 (z-t_1 (t_2+z-1))^2}\frac{dt_1}{2\pi i}\frac{dt_2}{2\pi i}\,.
\fe  
The contour integrals are performed as follows. We first perform the contour integral over $t_1$ around the pole at $t_1=z/t_2$.  This leaves a contour integral over $t_2$, for which the relevant poles are at $t (t_2-1) (t_2-z)+ t_2 (z-1)=0$.   The resulting residues at these poles lead to the final expression for \eqref{eq:examp1}, which is given by
\ie
\!\! \frac{3 t z \left(t^4 (z-1)^3-2 t^3 (z-1)^2 (z+1)+14 t^2 (z-1) z+2 t \left(z^3-5 z^2-5 z+1\right)-z^3-3 z^2+3
   z+1\right)}{2 (z-1)^{3\over 2} \left[ (t-1)^2 z-(t+1)^2 \right]^{7\over 2}}\,.
\fe
Similarly, one can determine  all the other contributions in \eqref{eq;so} to the generating function $\cB^1_{SO}(z;t)$. We have also rederived the function  $\cB_{SU}(z;t)$ using this method.

	\bibliographystyle{ssg}
	\bibliography{genfun-ref}

\end{document}